%% file: HiPC-abdulah.tex
\documentclass[conference]{IEEEtran}
\IEEEoverridecommandlockouts

\usepackage{amsmath,amssymb,amsfonts}
\usepackage{graphicx}
\usepackage{xcolor, soul}
\sethlcolor{green}

\usepackage{booktabs} 
\usepackage{caption,algpseudocode,algorithm,setspace}
\usepackage{scalerel,stackengine}
\usepackage{subfigure}
\usepackage{graphicx}
\usepackage{grffile}
\usepackage{tabu}
\usepackage{textcomp}
\usepackage{caption,algpseudocode,algorithm,setspace}
\usepackage{booktabs}   

\usepackage{graphicx}
\usepackage{subfigure}
\usepackage{amsfonts}
\usepackage{textcomp}
\usepackage{caption}
\usepackage{url}
\usepackage{amsmath}
\usepackage{breqn}
\usepackage{mathtools}
\usepackage{color}
\usepackage{flushend}
\usepackage{caption,algpseudocode,algorithm,setspace}
\usepackage{scalerel,stackengine}

\usepackage{listings}
\usepackage{balance}

\usepackage{xspace}

\newcommand{\starpu}{\textsc{StarPU}\xspace}

\newcommand{\exageostat}{\textsc{ExaGeoStat}\xspace}

\usepackage[textwidth=125pt,textsize=small,disable]{todonotes}

\usepackage{booktabs} 
\usepackage{listings}
\def\BibTeX{{\rm B\kern-.05em{\sc i\kern-.025em b}\kern-.08em
    T\kern-.1667em\lower.7ex\hbox{E}\kern-.125emX}}
    
\begin{document}

\title{Geostatistical Modeling and Prediction\\ Using Mixed-Precision Tile Cholesky Factorization
}

\author{\IEEEauthorblockN{Sameh Abdulah, Hatem Ltaief, Ying Sun, Marc G. Genton, and David E. Keyes}
\IEEEauthorblockA{\textit{Computer, Electrical, and Mathematical 
  Sciences and Engineering Division} \\
\textit{King Abdullah University of Science Technology}\\
Thuwal, 23955-6900, Saudi Arabia \\
sameh.abdulah@kaust.edu.sa, hatem.ltaief@kaust.edu.sa, ying.sun@kaust.edu.sa,\\ marc.genton@kaust.edu.sa,
and david.keyes@kaust.edu.sa}
}

\maketitle

\begin{abstract}
Geostatistics represents one of the most challenging classes of scientific applications due to the desire to incorporate
an ever increasing number of geospatial locations to accurately model and predict environmental phenomena.
For example, the evaluation of the Gaussian log-likelihood function, which constitutes the main computational phase,
involves solving systems of linear equations with a large dense symmetric and positive definite covariance matrix.
Cholesky, the standard algorithm, requires $O(n^3)$ floating point operators and has an $O(n^2)$ memory footprint,
where n is the number of geographical locations.  Here, we present a mixed-precision tile algorithm to accelerate the
Cholesky factorization during the log-likelihood function evaluation. Under an appropriate ordering, it operates
with double-precision arithmetic on tiles around the diagonal, while reducing to single-precision arithmetic for tiles
sufficiently far off. This translates into an improvement of the performance  without any deterioration of the numerical accuracy
of the application. We rely on the StarPU dynamic runtime system to schedule the tasks and to overlap them with
data movement. To assess the performance and the accuracy of the proposed mixed-precision algorithm, we use
synthetic and real datasets on various shared and distributed-memory systems possibly equipped with hardware accelerators.
We compare our mixed-precision Cholesky factorization against the double-precision reference implementation as
well as an independent block approximation method. We obtain an average of 1.6X performance speedup on massively
parallel architectures, while maintaining the accuracy necessary for modeling and prediction.
\end{abstract}



\input{./text/intro.tex}

\input{./text/related.tex}

\input{./text/contrib.tex}

\input{./text/pb.tex}

\input{./text/bg.tex}

\input{./text/proposed.tex}

\input{./text/impl.tex}

\input{./text/perf.tex}

\input{./text/summary.tex}

\input{./text/ack.tex}

\balance
\bibliographystyle{IEEEtran}
\bibliography{./bib/sample-bibliography,./bib/references}

\clearpage
\listoftodos[Notes]

\end{document}

%% file: text/intro.tex
\section{Introduction}
\label{sec:intro}
The rise of mixed-precision algorithmic developments in the scientific
community coincides with the advent of machine learning techniques
for performing analytics on big data problems. Because the convergence 
between HPC and big data~\cite{bdec-report} is now at the forefront of research
innovations for the digital world, e.g., healthcare, security,
and climate/weather modeling, hardware vendors have tremendously
invested in designing chips during the last decade with an emphasis 
on further provisioning low precision floating-point units~\cite{tensorcores,googletpu}.
This computing paradigm shift has mobilized researchers in identifying opportunities
within their legacy numerical algorithms to exploit such hardware features.
The main idea consists in determining which computational 
phases within an algorithm are resilient to a lower precision, while
ultimately maintaining the required level of accuracy for the final solution.
The availability of off-the-shelves hardware with effective support of low precision
floating-point arithmetics has further democratized mixed-precision algorithms.
This hardware/algorithm synergism has demonstrated to be a game-changing approach
for solving some of the most challenging scientific problems~\cite{gb-gwas-sc2019}.

Given this fertile landscape, we propose to study geostatistics
modeling and prediction using mixed-precision
algorithms. Rather than using first principles physics approaches,
geostatistics may represent a plausible alternative
to accurately model and predict environmental phenomena given the availability of measurements at a high number of geospatial locations. One of the main computational
phases necessitates the evaluation of the Gaussian log-likelihood function,
which translates into iteratively solving a number of large systems of linear
equations. The large dense symmetric and positive definite 
covariance matrix can be processed with the Cholesky factorization
with a cubic algorithmic complexity as the number of geographical locations 
with measurements grow. There are several methodologies to reduce this
intractable complexity, e.g., dimension-reducing PCA approaches, 
independent block approximation method~\cite{stein2014limitations},
or low-rank approximations~\cite{ Hackbusch1999}, to name a few. 
These methods are thoroughly described and evaluated in~\cite{sun2012geostatistics}.

In this paper, we present a new mixed-precision algorithm to 
accelerate the Cholesky factorization during the log-likelihood function evaluation
in the context of environmental applications.
Based on tile algorithms~\cite{Agullo_2009_jpcs}, the resulting Cholesky factorization
takes advantage of the covariance matrix structure, which retains the most
significant information around the diagonal of the matrix. Instead of
completely annihilating the off-diagonal contributions engendering a possible
loss of accuracy, we operate the computation of close-to-diagonal tiles 
in  double-precision accuracy while switching to single-precision
accuracy for the remaining far off-diagonal tiles. Insight into the application data
sparsity is paramount to take into account, before moving
forward with mixed-precision algorithms. While mixed-precision algorithmic
optimization  translates into performance gains, it is critical
to validate the statistical parameter estimators, which drive the modeling
and the prediction phases for climate and weather applications.

To cope with the heterogeneity of the mixed-precision workloads, we rely on
the \starpu dynamic runtime system~\cite{augonnet2011starpu} to schedule the various tasks 
onto available resources, while overlapping the expensive data movement with useful
computations. We assess the numerical accuracy and the performance of the new
mixed-precision Gaussian log-likelihood function using synthetic and real datasets
on a myriad of shared and distributed-memory systems possibly equipped with 
GPU hardware accelerators. We report the performance of our novel mixed-precision
Cholesky factorization against the double-precision reference implementation as well
as an independent block approximation method. Our benchmarking results reveal a significant improvement
of the performance speedup on these massively parallel architectures,
with an average of 1.6X performance speedup on these massively parallel architectures, 
while maintaining the necessary accuracy for modeling and prediction purposes. 
This latest algorithmic optimization further extends the features of our
software framework \exageostat, which packages high performance implementations of
algorithmic adaptations for large-scale environmental applications.
To the best of our knowledge, this is the first implementation of a mixed-precision Cholesky
factorization applied on a tile basis, using a non-iterative approach. Although this paper focuses on  applications
for geostatistics, the Cholesky factorization is a pivotal matrix operation for several other big data applications.

The remainder of the paper is organized as follows.
Section~\ref{sec:related} presents related work and Section~\ref{sec:contrib}
lists our research contributions. Section~\ref{sec:pb} provides the necessary background
for geostatistical applications. Section~\ref{sec:bg} recalls the current algorithmic features
of our \exageostat software framework. Section~\ref{sec:mixed} introduces our novel
tile Cholesky factorization using mixed-precision techniques. We detail its parallel
implementation in Section~\ref{sec:impl}. Section~\ref{sec:perf} reports the accuracy and the performance
results of our mixed-precision tile Cholesky factorization on various hardware architectures 
using synthetic and real datasets. Section~\ref{sec:summary} summarizes our contributions
from the paper and highlights future work.

%% file: text/related.tex
\section{Related Work}
\label{sec:related}
Several approximation techniques have been proposed in the literature to reduce the
arithmetic complexity and memory footprint in large-scale problems. 
In ~\cite{banerjee2008gaussian}, Gaussian Predictive Processes (GPP) was proposed
to reduce the dimensions of large scale covariance matrices. This reduction is achieved by projecting the original problem into
a subspace at a certain set of locations. However, this method usually underestimates
the variance parameter~\cite{sun2012geostatistics}. Another approach based on
fixed-rank kriging has been proposed by~\cite{cressie2008fixed}. 
This approach uses a spatial mixed effects model for the spatial problem and proposes
fixed rank kriging using a set of non-stationary covariance functions.
In~\cite{kaufman2008covariance}, a covariance tapering approach has been proposed by
converting the given dense covariance to a sparse matrix. 
The sparse matrix is generated by ignoring the large distances and set them to zero.
In this case, sparse matrices algorithms can be used for fast computation.
Other methods such as  Kalman filtering~\cite{sinopoli2004kalman}, 
moving averages~\cite{ver2004flexible},
and low-rank splines~\cite{kim2004smoothing} have been proposed to approximate
the covariance matrix by reducing the problem dimension.

Hierarchical matrices ($\mathcal{H}$-matrices) are widely used to accommodate the large covariance
matrices dimension by applying a low-rank approximation to the off-diagonal
matrix~\cite{litvinenko2019likelihood}.
Different data approximation techniques based on $\mathcal{H}$-matrices have been proposed
in literature such as Hierarchically Off-Diagonal Low-Rank (HODLR)~\cite{aminfar2016fast},
Hierarchically Semi-Separable (HSS)~\cite{ghysels2016efficient},
$\mathcal{H}^2$-matrices~\cite{borm2016approximation,sushnikova2016preconditioners},
and Block/Tile Low-Rank (BLR/TLR)~\cite{pichon2017sparse, akbudak2018exploiting}.

Most of the existing studies about mixed-precision in climate and weather applications are
related to the analysis of the effect of applying mixed-precision to such applications. For instance,
in~\cite{duben2014use}, the authors show that low precision arithmetic coming
from faulty hardware has a negligible effect on the overall accuracy of weather
and climate prediction. The study covers only the analysis part of the mixed-precision
impact on such applications. Another study by~\cite{thornes2017use} examines how a mixture of single- and half-
precision could be useful in the case of weather
and climate applications. In~\cite{gan2013accelerating}, a hybrid CPU-FPGA algorithm
is proposed with mixed-precision support to compute the upwind stencil for the
global shallow water equations with magnificent speedup, compared to pure CPU and hybrid CPU-GPU systems.

Mixed-precision iterative refinement approaches have been studied for solving 
dense linear system of equations~\cite{Buttari-iterref} using single and double-precision
arithmetics. A new mixed
precision iterative refinement approach~\cite{Haidar-iterref} has shown a significant improvement of the performance 
(speedup factor up to four) using multiple precisions, i.e., 16-bit, 32-bit, 
and 64-bit precision arithmetics for the dominant GEMM kernel, on NVIDIA V100 GPUs.
These mixed-precision approaches use a unique precision arithmetic for the
Cholesky factorization and subsequently, iterate using multiple precisions to refine the solution.
There are, however, numerical restrictions depending on the number of matrix conditions.
There are also recent works toward democratizing half-precision arithmetics for
climate applications for accelerating DL workloads~\cite{Kurth-dlclimate}
achieving tremendous performance speedups.

Last but not least, the presented mixed-precision Cholesky factorization may accelerate
scientific applications beyond the one studied herein. It may be applied
to computational astronomy applications, which consist in removing the impact of
the atmospheric turbulence on the distorted light from the remote galaxies and captured
by ground-based telescopes~\cite{astronomy}. Moreover, Calculating the electronic structure
of molecules in material sciences may translate into solving the Schr\"{o}dinger equation,
using a generalized symmetric eigenvalue decomposition. The mixed-precision method may be applied during the first computational phase of the eigensolver to approximate
the low interactions between distant electrons~\cite{gevp}.


In this paper, we adopt the covariance tapering approach. The default approach
consists in ignoring the correlations between the remote spatial locations separated
with a predetermined distance by setting them to zero.
Instead, we use flexible and adaptive mixed-precision algorithm to reduce the
precision accuracy of these correlation values. We then launch the new high performance Cholesky
factorization operating on mixed-precision tile data structures, which represents the main engine 
driving the maximum likelihood estimation (MLE). This may ultimately permit to
achieve better numerical accuracy than the covariance tapering approach
for the prediction.

%% file: text/contrib.tex
\section{Contributions}
\label{sec:contrib}
Our contributions can be summarized as follows: (1) we propose a novel mixed-precision
Cholesky factorization algorithm to accelerate the maximum likelihood 
evaluation (MLE); (2) we apply the new algorithm to the climate weather modeling and predictions problems by extending the
existing {\em ExaGeoStat} software with the new mixed-precision Cholesky algorithm; (3) we conduct
a set of experiments to validate the accuracy of the proposed algorithm and to show its 
ability to satisfy the accuracy requirements of the climate and weather applications; (4)
we provide a quantitative performance analysis to assess the performance of the proposed
algorithm on heterogenous shared-memory, and distributed-memory environments.

%% file: text/pb.tex
\section{Geostatistics Applications}
\label{sec:pb}

\subsection{Background}

Climate and environmental data usually include
a large number of measurements distributed
regularly or irregularly across a given geographical
region. Each location is associated with a single value
of climate or weather variable such as temperature,
precipitation, wind speed, air pressure, etc. These data can be modeled
as a realization from a  Gaussian spatial random field
when considering geostatistics applications.
Specifically, for a given geographical region $\Bbb{R}^d$, let $n$
represents the number of available spatial locations
 from ${\bf s}_1$ to ${\bf s}_n$, and let ${\bf Z}=\{Z({\bf s}_1),\ldots,Z({\bf s}_n)\}^\top$ be a realization 
of a Gaussian random field $Z({\bf s})$,
i.e., measurements, at those $n$ locations. 
Assume the random field $Z({\bf s})$ 
has a mean zero and stationary parametric
covariance function 
$C({\bf h};{\boldsymbol \theta})=\mbox{cov}\{Z({\bf s}),Z({\bf s}+{\bf h})\}$,
where ${\bf h}\in\Bbb{R}^d$ is a spatial lag
vector and ${\boldsymbol \theta}\in\Bbb{R}^q$ is
an unknown parameter vector.

\subsection{Mat\'{e}rn Covariance Function}
The Mat\'{e}rn class of covariance functions is a
 generic form that is used to construct the covariance 
matrix ${\boldsymbol \Sigma}({\boldsymbol \theta})$ in geostatistics applications. The Mat\'{e}rn
 function is defined as,

\begin{equation}
	C(r;{\boldsymbol \theta})=\frac{\theta_1}{2^{\theta_3-1}\Gamma(\theta_3)}\left(\frac{r}{\theta_2}\right)^{\theta_3} {\mathcal K}_{\theta_3}\left(\frac{r}{\theta_2}\right),
		\label{eq:materncov}
\end{equation}
where ${\boldsymbol \Sigma}({\boldsymbol \theta})$ is  a symmetric positive definite covariance matrix  with
entries ${\boldsymbol \Sigma_{ij}}=C({\bf s}_i-{\bf s}_j;{\boldsymbol \theta})$, $i,j=1,\ldots,n$, and
 ${\boldsymbol \theta}=(\theta_1,\theta_2,\theta_3)^\top$ is the model parameter vector.
  Here, $\theta_1>0$ is the variance parameter,
$\theta_2>0$ is a spatial range parameter that
 measures how quickly the correlation of the random 
field decays with distance, and $\theta_3>0$ controls
 the smoothness of the 
random field, with larger values of $\theta_3$ corresponding
 to smoother fields. 
Here, $r=\|{\bf s}-{\bf s}'\|$ is the distance between any two spatial
 locations, ${\bf s}$ and ${\bf s}'$. This distance
can simply be calculated using Euclidean Distance (ED)
 metric in small geographic areas. For more accurate
estimation in large areas, the Great-Circle Distance (GCD) 
metric should be more appropriate~\cite{rick1999deriving, abdulah2018exageostat},
${\displaystyle \operatorname {hav} \left({\frac {d}{r}}\right)=\operatorname {hav} (\varphi _{2}-\varphi _{1})+\cos(\varphi _{1})\cos(\varphi _{2})\operatorname {hav} (\lambda _{2}-\lambda _{1})}$
, where hav is the haversine function, 
$\operatorname {hav} (d )=\sin ^{2}\left({\frac {d }{2}}\right)={\frac {1-\cos(d )}{2}}$;
$d$ is  the central angle between any two points on a sphere,   
$\varphi_1$ and $\varphi_2$ are the latitudes in
 radians of locations $L_1$ and 
$L_2$, respectively, and $\lambda _{1}$ and $\lambda _{2}$ 
are longitudes.


\subsection{Maximum Likelihood Estimation (MLE)}

Statistical inference about $\boldsymbol \theta$ is
 often based on the Gaussian
log-likelihood function,
\begin{equation}
	\label{eq:likeli}
	\ell({\boldsymbol \theta})=-\frac{n}{2}\log(2\pi) - \frac{1}{2}\log |{{\boldsymbol \Sigma}({\boldsymbol \theta})}|-\frac{1}{2}{\bf Z}^\top {\boldsymbol \Sigma}({\boldsymbol \theta})^{-1}{\bf Z},
\end{equation}
where ${\boldsymbol \theta} = ( \theta_1, \theta_2, \theta_3)$ 
and the maximum likelihood estimator of
${\boldsymbol \theta}$ is the value $\widehat{\boldsymbol \theta}$
 that maximizes (\ref{eq:likeli}). 

Equation (\ref{eq:likeli}) usually requires the optimization of three model
parameters, i.e., variance, range, and smoothness; this involves a large number of iterations.
The above equation can be simplified to limit the number of these iterations by reducing the number of optimized parameters.
 This
modification requires the optimization of  two parameters only, i.e., $ \theta_2$ and $\theta_3$, and considers
$\theta_1$ as a multiplicative scale parameter that can be
computed directly from the optimized parameters  ${\boldsymbol {\tilde \theta}} = ( \theta_2, \theta_3)$. In this case, equation (\ref{eq:likeli}) can be represented as,

\begin{dmath}
	\label{eq:likeli2}
	\ell({\boldsymbol {\tilde \theta}}, {\theta_1}^{opt})=-\frac{n}{2}\log(2\pi) - \frac{n}{2}+ \frac{n}{2} log (n) -  \frac{1}{2}\log |{{\boldsymbol { \tilde{\Sigma}}}({\boldsymbol \theta})}|-\frac{1}{2}{\bf Z}^\top {\boldsymbol{ \tilde{\Sigma}}}({\boldsymbol \theta})^{-1}{\bf Z}.  
\end{dmath}

The optimized  parameter $\theta_1$ can be obtained at the
end of the optimization problem with 
	${\theta_1}^{opt}=\frac{1}{n}{\bf Z}^\top {\boldsymbol{ \tilde{\Sigma}}}({\boldsymbol \theta})^{-1}{\bf Z} $  where ${\boldsymbol{ \tilde{\Sigma}}}$ is the covariance matrix generated using $\boldsymbol {\tilde \theta}$.

A large-scale evaluation of MLE  is a prohibitively expensive operation
due to the necessary  floating-point operations and memory.
Thus, the evaluation of equation (\ref{eq:likeli2}) is challenging because of the 
linear solver and log-determinant involving the $n$-by-$n$
dense and unstructured covariance matrix
${\boldsymbol \Sigma}({\boldsymbol \theta})$  and requires
 ${\mathcal O}(n^3)$ floating-point operations
on ${\mathcal O}(n^2)$ memory. In real applications, some
 datasets  can contain millions 
of locations which requires a huge memory footprint that
can reach to tens or even hundreds of terabytes. In this	
paper, we give one example of a real dataset from the 
Middle East region with $\sim1M$ locations.

%

%% file: text/bg.tex
\section{The \exageostat Software }
\label{sec:bg}

\begin{figure*}
	\centering
	\subfigure[Full Double-Precision (DP).]{
		\label{fig:exa-exact}
	\includegraphics[width=0.2\linewidth]{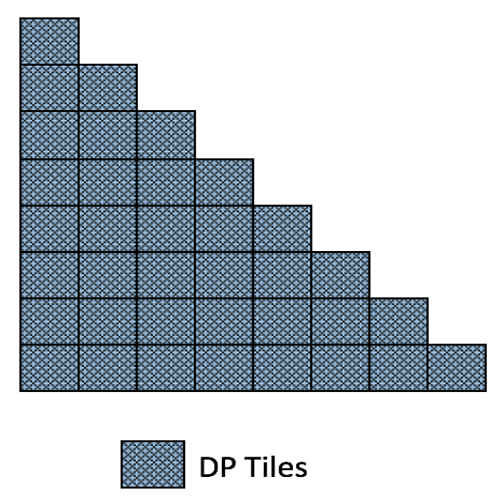}}
		\hspace{0.3 cm}
		\subfigure[Diagonal Super-Tile (DST).]{
		\label{fig:exa-dst}
	\includegraphics[width=0.2\linewidth]{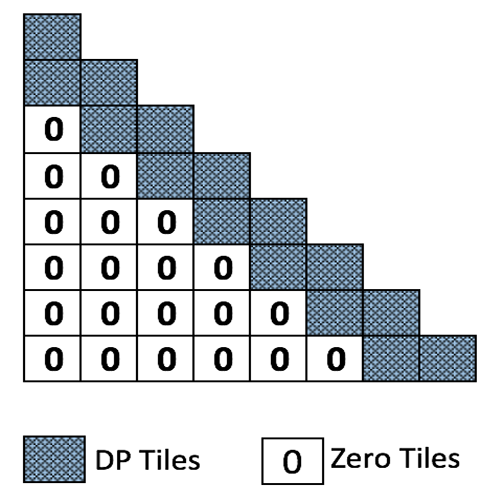}}
		\hspace{0.3 cm}
		\subfigure[Tile Low-Rank (TLR).]{
		\label{fig:exa-tlr}
	\includegraphics[width=0.205\linewidth]{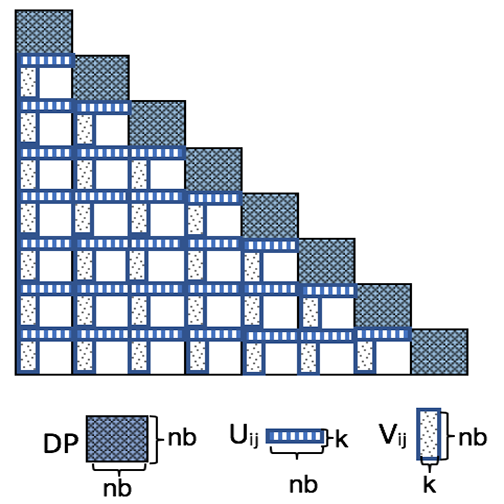}}
		\hspace{0.3 cm}
		\subfigure[Mixed-Precision (MP).]{
		\label{fig:mpexample}
	\includegraphics[width=0.2\linewidth]{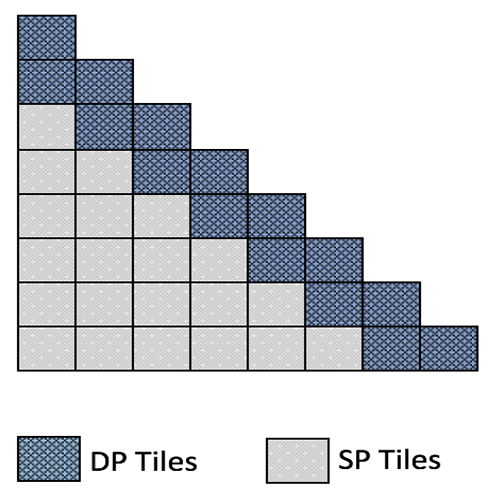}}
	
	\caption{Four ways of computing the Cholesky-based MLE in \exageostat.}
	\label{fig:exageostat}
\end{figure*}

The \exageostat software  presents a solution for the exascale computing
used in geostatistical modeling and prediction based on the Cholesky factorization
to drive the maximum likelihood estimation (MLE).
\exageostat currently supports three ways of doing Cholesky factorization to perform the MLE operation,
as depicted in Fig.~\ref{fig:exa-exact}, Fig.~\ref{fig:exa-dst}, and Fig.~\ref{fig:exa-tlr}.
The factorization variant depends on the approximation type applied to the given covariance matrix, i.e.,
full double-precision dense structure, IndepeNDent blocks/Diagonal Super-Tile
(IND/DST) structure, or Tile Low-Rank (TLR) structure. The following subsections
give a brief background on each method.


\subsection{Dense Tile Cholesky Factorization}

Fig.~\ref{fig:exa-exact} gives an example of dense
computation on a tile-based matrix.
All the tiles are represented in double-precision 
and a set of Level 3 BLAS routines are applied to perform the Cholesky factorization of
a given $n \times n$  symmetric, positive-definite matrix.
In this paper, we assume that the factorization of $\mathbf{ A}$ is presented by $\mathbf{LL}^\top$
and $\mathbf{L}$ is an $n \times n$ real lower triangular matrix with positive
diagonal elements.

\subsection{Diagonal Super-Tile (DST) Cholesky Factorization}
Covariance tapering is commonly used to 
approximate the covariance matrix  from  geostatistics applications
by ignoring the correlation with the very far 
locations~\cite{kaufman2008covariance, stein2013statistical}. 
In~\cite{huang2018hierarchical}, 
a tapering algorithm is proposed by setting
the correlation between any two far locations $i$ and $j$ equals
to zero. This method has been called in statistics as
{\em Independent blocks method (IND)}. The blocks represent 
different parts of the whole geographical area with maximum
dimension equals to the maximum distance between two
correlated locations. In~\cite{abdulah2018exageostat}, 
we have implemented the IND method by depending on
the diagonal elements where the maximum distance can be
represented by the number of in-used diagonal tiles. We have called
this implementation as the {\em Diagonal Super-Tile (DST)} method. 
Fig.~\ref{fig:exa-dst} gives an example of the DST method 
where two diagonal tiles are represented in full
precision and the other tiles are set to zeros.

Although the IND/DST approach seems impractical in many cases, since it
ignores the existing relation between some of the spatial locations; in some cases, 
an IND/DST approach can be better than sophisticated likelihood
approximation methods such as low-rank methods~\cite{stein2014limitations}, for instance,
when observations are dense enough and with low noise effects on the correlations
between existing spatial locations. In general,
an IND/DST approach can be used
to reduce the space and computing complexity of dense computations,
as long as it achieves the required accuracy.

\subsection{Tile Low-Rank (TLR) Cholesky Factorization}
Low-rank approximation is a common way to approximate geostatistics covariance matrices~\cite{Hackbusch1999}.
 In~\cite{abdulah2018parallel},
we have leveraged \exageostat to support Tile Low-Rank (TLR) approximation. The TLR 
solution depends on exploiting the data sparsity of the given covariance matrix by compressing 
the off-diagonal tiles up to a certain accuracy level.  In TLR approximation, 
the diagonal tiles are kept as dense while the Singular Value
Decomposition (SVD) technique is used to approximate the off-diagonal tiles.
Each off-diagonal tile is represented by two matrices $\mathbf{U}$ and $\mathbf{V}$ representing
the most significant $k$ singular values and their associated left
and right singular vectors, respectively. $k$ is the actual rank of each tile. In general, variables
ranks are expected across different matrix tiles. In the case of low accuracy level, small
ranks lead to low memory footprint while in the case of high accuracy level, large ranks
lead to high memory footprint.

Fig.~\ref{fig:exa-tlr} gives an example of applying TLR
to a given covariance matrix. Each off-diagonal tile $(i, j)$ is represented by  the product
of $U_{ij}$, with size $k \times nb$ and $V_{ij}$ with size $nb \times k$, where $nb$ represents
the tile size. $nb$ should be tuned in different hardware architectures
to obtain the best performance since it corresponds to the trade-off
between arithmetic intensity and degree of parallelism.

%% file: text/proposed.tex
\section{Mixed-precision Cholesky Factorization}
\label{sec:mixed}

\begin{figure*}
	\centering
	\includegraphics[width=17 cm]{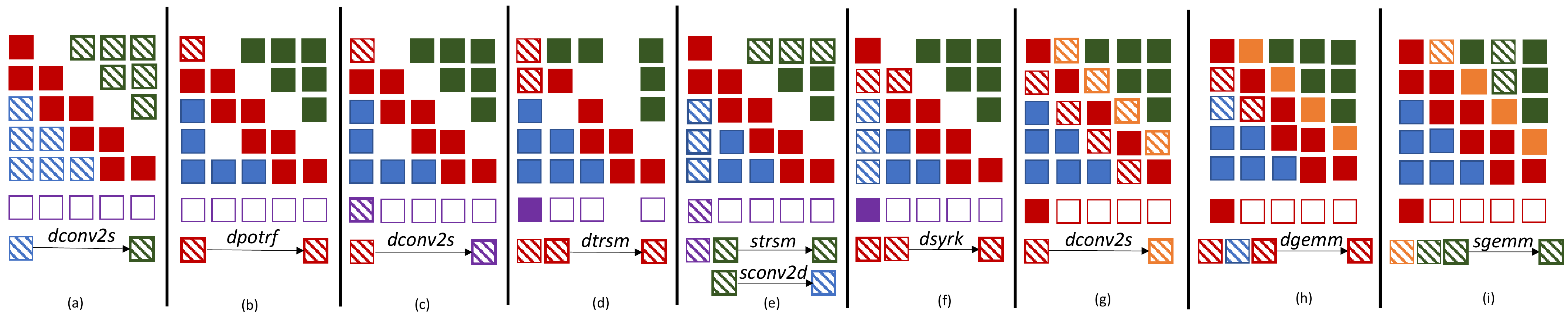}
	\caption{
	Unrolling the first outer loop iteration of
	Algorithm~\ref{alg:mpmle} with a
	$5 \times 5$ tile matrix size and a diagonal thickness set to 2.}
	\label{fig:impl_example}
\end{figure*}

The Cholesky factorization represents one of the most time-consuming operations,
when computing the likelihood function given by equation~\ref{eq:likeli}.
In geostatistics applications, a double-precision arithmetic
is usually required to satisfy the desired accuracy
level~\cite{abdulah2018exageostat, abdulah2018parallel}.
 Assuming an appropriate ordering of the spatial locations, the covariance matrix
 from those applications has the most valuable information around the diagonal elements
of the matrix. The closest locations to a certain location are more correlated to this location, in comparison with far locations.  Thus, some methods have proposed to ignore
the relation between far locations by ignoring their impact on the covariance matrix, i.e., 
set to zero~\cite{stein2013statistical}.

In this paper, we propose a mixed-precision  Cholesky factorization algorithm
based on the tile-based Cholesky factorization algorithm. 
This algorithm aims at keeping valuable
information around the diagonal elements by using a double floating
precision in the diagonal elements, and at the same time, representing
the off-diagonal elements in single floating precision
to accelerate the algorithm execution time while preserving the accuracy
required by the application. However, applying a mixed floating precision
is a tricky process, since the computation of each tile depends on
 other tiles with different data format representation.
 
 To compute
the Cholesky factorization of a given dense matrix, only the upper
or the lower part of the matrix is used to store the output of the operation, i.e., $\mathbf{A}=\mathbf{LL}^\top$. Our proposed
implementation uses the other part of the matrix to store
the single-precision representation of the matrix tiles, and recalls them when needed. 
For the diagonal tiles, a tile vector of size
$n \times nb$ is required to store the single-precision representation of these tiles. 

%% file: text/impl.tex
\section{Implementation Details}
\label{sec:impl}

Algorithm~\ref{alg:mpmle} shows the proposed mixed-precision algorithm. 
A symmetric positive-definite matrix $\bf A$, and the required accuracy level $diag\_thick$
are the two required inputs to the algorithm. The accuracy level represents the number
of the diagonal tiles that are represented as double-precision while the
remaining off-diagonal tiles are represented in single-precision. Fig.~\ref{fig:mpexample} 
shows an example of a mixed-precision data formatting with diagonal thickness = 2 while 
Fig.~\ref{fig:impl_example} gives a step-by-step example of Algorithm~\ref{alg:mpmle} using a
$5 \times 5$ matrix and two double-precision diagonal tiles.

\begin{algorithm} [htp]
\setstretch{1.1}
\footnotesize
\caption{Mixed-precision tile Cholesky factorization (lower triangular case)}
\label{alg:mpmle}
\begin{algorithmic}[1]
\State Input:  \texttt{A symmetric positive definite matrix $\bf A$, and the accuracy level $ diag\_thick$}.

	\For{\texttt{i=1,2,...,p}}
				 \For{\texttt{j=i+$diag\_thick$,...,p}}
				  		\State $\bf{A}_{ji}$ $\leftarrow$ dlag2s($\bf{A}_{ij}$)
				\EndFor	
	\EndFor		 
\For{\texttt{k=1,2,...,p}}
		 \State \texttt{$\bf{A}_{kk}$ $\leftarrow$ dpotrf($\bf{A}_{kk}$)}
		 				  		\State \texttt{$\bf{tmp}_{0k}$ $\leftarrow$ dlag2s($\bf{A}_{kk}$)}
	\For{\texttt{i=k+1,...,p}}
	 		\If {\texttt{|m-k| < $ diag\_thick$}   	}   
				 \State \texttt{$\bf{A}_{ik}$ $\leftarrow$ dtrsm($\bf{A}_{kk}$,$\bf{A}_{ik}$)}
			\Else
		\State \texttt{$\bf{A}_{ki}$ $\leftarrow$ strsm($\bf{tmp}_{0k}$,$\bf{A}_{ki}$)}
       \State \texttt{$\bf{A}_{ik}$ $\leftarrow$ sconv2d($\bf{A}_{ki}$)}
			 		 \EndIf

				\EndFor	
		\For{\texttt{j=k+1,...,p}}
				\State \texttt{$\bf{A}_{jj}$ $\leftarrow$ dsyrk($\bf{A}_{jk}$,$\bf{A}_{jj}$)}
		 		\If {\texttt{|k-j| < $ diag\_thick$}  } 
				\State \texttt{$\bf{A}_{kj}$ $\leftarrow$ dconv2s($\bf{A}_{jk}$)}
			 		 \EndIf

				 \For{\texttt{i=j+1,...,r}}
				 
				 	 		\If {\texttt{|m-k| < $ diag\_thick$}   	  }
					 \State \texttt{$\bf{A}_{ij}$ $\leftarrow$ dgemm($\bf{A}_{ik}$,$\bf{A}_{jk}$,$\bf{A}_{ij}$)}
			\Else
					 \State \texttt{$\bf{A}_{ji}$ $\leftarrow$ sgemm($\bf{A}_{ki}$,$\bf{A}_{kj}$,$\bf{A}_{ji}$)}

			 		 \EndIf
						 
				\EndFor	
		\EndFor	
\EndFor		
\end{algorithmic}
\end{algorithm}

The algorithm begins by converting all the off-diagonal tiles
to a single-precision format using $dconv2s$ kernel, and
stores the results to the upper triangular part of the matrix (i.e., assuming lower
triangular Cholesky factorization is applied) (lines 2-6). The conversion
includes a transpose operation to the selected tile $(i, j)$. This step is also
shown by Fig.~\ref{fig:impl_example}(a). In line 8, a double-precision Cholesky
factorization is applied to the diagonal tile ($k$, $k$) (Fig.~\ref{fig:impl_example}(b)).
This diagonal tile is used to perform some single-precision operations on
the tiles at the same column $k$. Thus, a single-to-double conversion operation
$sconv2d$ is applied to the ($k$, $k$) tile, and the output is stored in a temporary
vector $tmp$ (line 9 and Fig.~\ref{fig:impl_example}(c)).
In lines 10-17, $trsm$ operation is applied to all the tiles of column $k$.
This operation can be a single or double-precision based on
the location of the target tile. The condition statement in line 11 is used to determine
the required precision for the $trsm$ operation. If the ($i$, $k$) tile is a diagonal tile
from the definition of the diagonal thickness, a $dtrsm$ operation is then
applied (Fig.~\ref{fig:impl_example}(d)), otherwise, a $strsm$ operation is applied
to the single-precision tiles of ($k$, $k$) stored at tile ($0$, $k$) in vector $tmp$ and the ($k$, $i$) 
tile (Fig.~\ref{fig:impl_example} (e)). In the case of a single-precision, the $sconv2d$ operation should
be applied to update the value of the double tile. In line 19, a $dsyrk$ operation is applied to the diagonal tile ($j$, $j$) 
(Fig.~\ref{fig:impl_example} (f)). In lines 20-21, all the double-precision
tiles except the diagonal tiles ($j$, $j$) are converted to single-precision
(Fig.~\ref{fig:impl_example} (g)). In lines 23-29, a $dgemm$ operation
is applied if the ($i$, $j$) tile is a double-precision tile (Fig.~\ref{fig:impl_example}(h)),
otherwise, a $sgemm$ operation is applied (Fig.~\ref{fig:impl_example}(i)).

%% file: text/perf.tex
	\section{Experimental Results}
	\label{sec:perf}
	This section presents the evaluation of the proposed mixed-precision
	 versus the full double-precision Cholesky algorithm. The evaluation involves assessing
	the performance of the MLE algorithm on heterogenous shared-memory,
	and distributed-memory systems.	 The  assessment of the accuracy involves the estimation of
	 the MLE model parameters	and the Prediction Mean Square Error (PMSE),
	 in the context of climate/weather applications  using both synthetic and real datasets.

	
	\subsection{Hardware and Software Platform}
	Our experiments were performed on various hardware architectures.
	For the  shared-memory systems  performance assessment, we used two Intel processors,
	 a 18-core dual-socket Intel Haswell chip and a 28-core dual-socket Intel Skylake
	chip. For the heterogeneous shared-memory 
	(CPU/GPU) systems  performance assessment, we used three Intel processors equipped
	 with different GPUs accelerators,
	a 14-core dual-socket Intel Broadwell chip equipped with a Nvidia Tesla K80 Kepler GPU.
	a 18-core dual-socket	Intel Haswell chip equipped with a Nvidia Tesla P100 Pascal GPU,
	and
	a 20-core dual-socket 	Intel Skylake chip equipped with  a Nvidia Tesla V100 Volta GPU,
		 For the distributed
         memory assessment, we used
	Shaheen-II, a Cray XC40 system with 6174 compute nodes based on dual-socket
	16-core Intel Haswell processors running at 2.3 GHz. Each node has a 128 GB
	 of DDR4 memory. The Shaheen-II has a total of  790 TB
	 of aggregate memory.
	
	Our code was
	compiled with gcc v5.5.0 and linked
	against the  Chameleon library\footnote{https://gitlab.inria.fr/solverstack/chameleon}
	with HWLOC v1.11.8, StarPU v1.2.4, Intel MKL 2018, GSL v2.4, CUDA v9.0,
	NLopt v2.4.2 optimization libraries, HDF5 v1.10.1, and NetCDF v4.5.0. 
	Our code will be integrated soon to the \exageostat software\footnote{https://github.com/ecrc/exageostat}.	
	
	
	\subsection{Experimental Testbed Datasets}
	A set of synthetic datasets and one real dataset
were used to evaluate the proposed algorithm.
	In the ensuing subsections, we provide more
	 details about these datasets.
	
	\subsubsection{Synthetic Datasets}
	\label{subsec:data_generator}
	\exageostat  provides an internal  data generator to
	simulate synthetic geostatistical data, based on the Mat\'{e}rn
	 covariance function. The synthetic data generation process includes
	 two main operations. First, a set of random 2D irregular spatial
	 locations were generated between $]0,1[$. Second, an initial parameter
	vector $\boldsymbol {\theta_0}$ was used to generate a set of $\boldsymbol Z$
	measurement vectors associated with the generated locations.  
 More details on the \exageostat data generator tool
	can be found in~\cite{abdulah2018exageostat}.

	\subsubsection{Real Dataset}
	\label{sec:real-dataset}
	we  also selected a real geostatistical dataset, the wind speed dataset,
	coming from the Middle East region to assess our proposed algorithm.
   Wind speed is a result of changing temperature through the air.	
   The movements of the air  from high-pressure to low-pressure layers, or vice versa,
   impact on the wind speed  measurements.
	The importance of wind speed, as one of the weather components, comes from the fact that
	it generally affects different activities related to
	both air and maritime transportation. 
	Additionally,
	various construction activities from airports to small houses 
	can be impacted by the speed and direction of the wind.

	The target dataset was generated using  WRF (WRF-ARW) software~\cite{skamarock2005description}.
	The  dataset was generated over the
	Arabian Peninsula in the Middle-East region.	Daily data are available for over 37 years; Each data file represents 24
	hours measurements of wind speed recorded hourly on 17
	different layers. In our case, we  picked up one dataset
	on September 1st, 2017 at time 00:00 AM on a 10-meter
	distance above the ground (i.e., layer 0). No special restriction
	is applied to the chosen data. We only select an example to
	show the effectiveness of our proposed mixed-precision method, but may
	easily consider extending the datasets.
	
\begin{figure}
	\centering
		\includegraphics[width=0.65\linewidth]{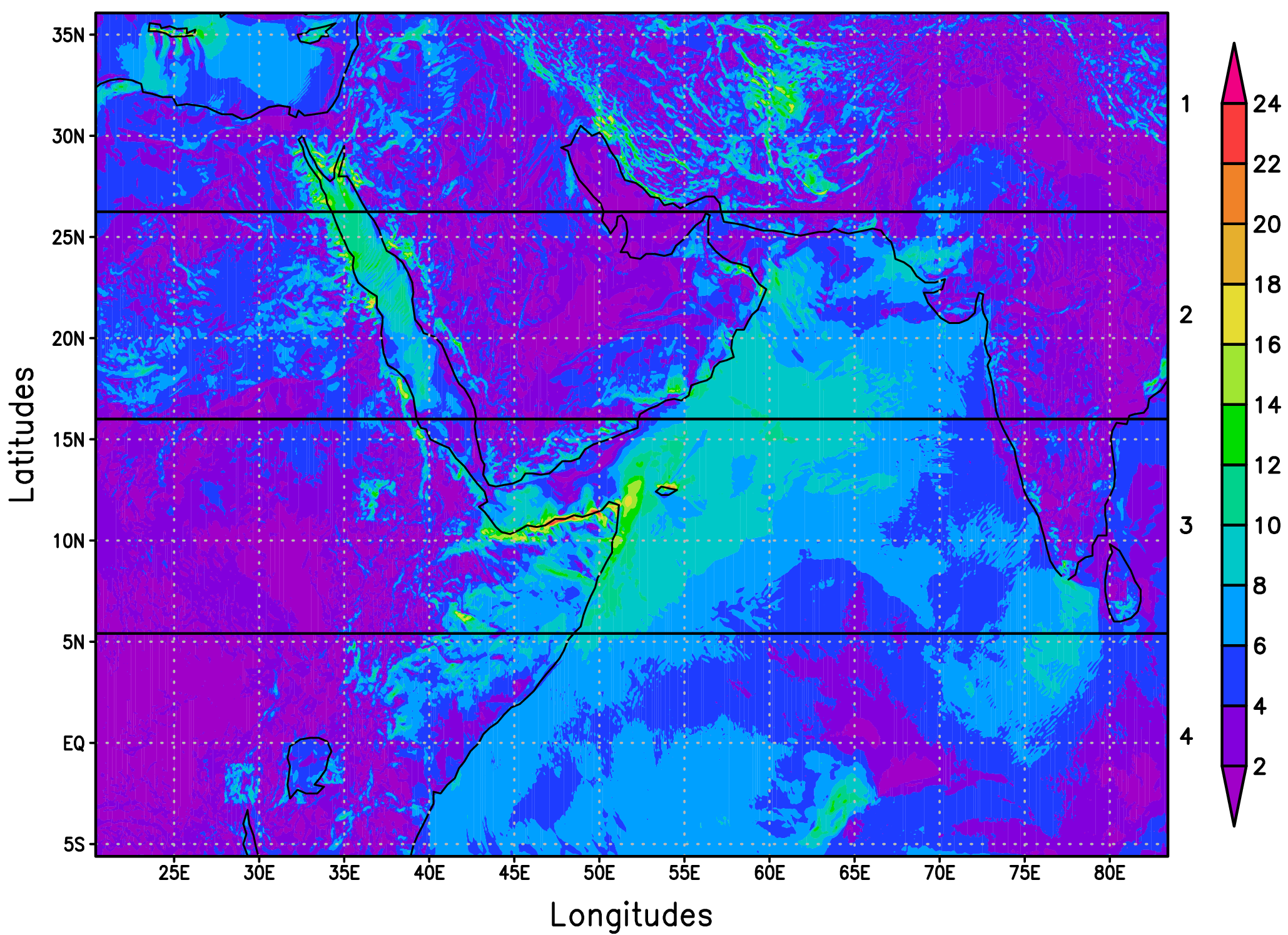}
		\caption{An example of climate/weather geostatistics real data from the Middle East (wind speed data).}
		\label{fig:wind-speed}
	\end{figure}

	Fig.~\ref{fig:wind-speed} shows the wind speed data where the locations
	were divided into four subregions (i.e., 1, 2, 3, and 4). Each region has approximately
	$250K$ locations. We choose to divide the dataset into four regions to avoid
	the non-stationarity exhibition,
	and to provide four different regions to properly assess the accuracy of  our method
	using different model parameters.

	\subsection{Mixed Precision  MLE Performance Evaluation}
	In this section, we provide a set of experiments to evaluate the performance of the proposed
	mixed-precision, in comparison with the full double-precision Cholesky algorithm 
	in the context of the MLE operation.
	We report the results on heterogenous shared-memory, and 
	distributed-memory systems.
	The reported time is the average time of one evaluation of the likelihood function. 
	All the plots display how the proposed mixed-precision algorithm outperforms
	the full double-precision algorithm 
	 on different architectures.
	All figures use different variants of the MLE algorithm. Here,
	we use {\em DP}  to represent the  pure double-precision arithmetic method, 
	and {\em DP($x\%$)-SP($y\%$)}  to represent the mixed-precision method where $x\%$ represents
	the amount of the  diagonal tiles that manipulates double-precision and 
 	$y\%$ is the amount of the  off-diagonal tiles that operates on single-precision.

	 \subsubsection{Homogeneous Shared-Memory Architectures}
	 \begin{figure}[!ht]
	 	\scriptsize
	 	\centering
	 \subfigure[Intel Haswell (36-core).]{
	 		\label{fig:time-haswell}
	 		\includegraphics[width=0.65\linewidth]{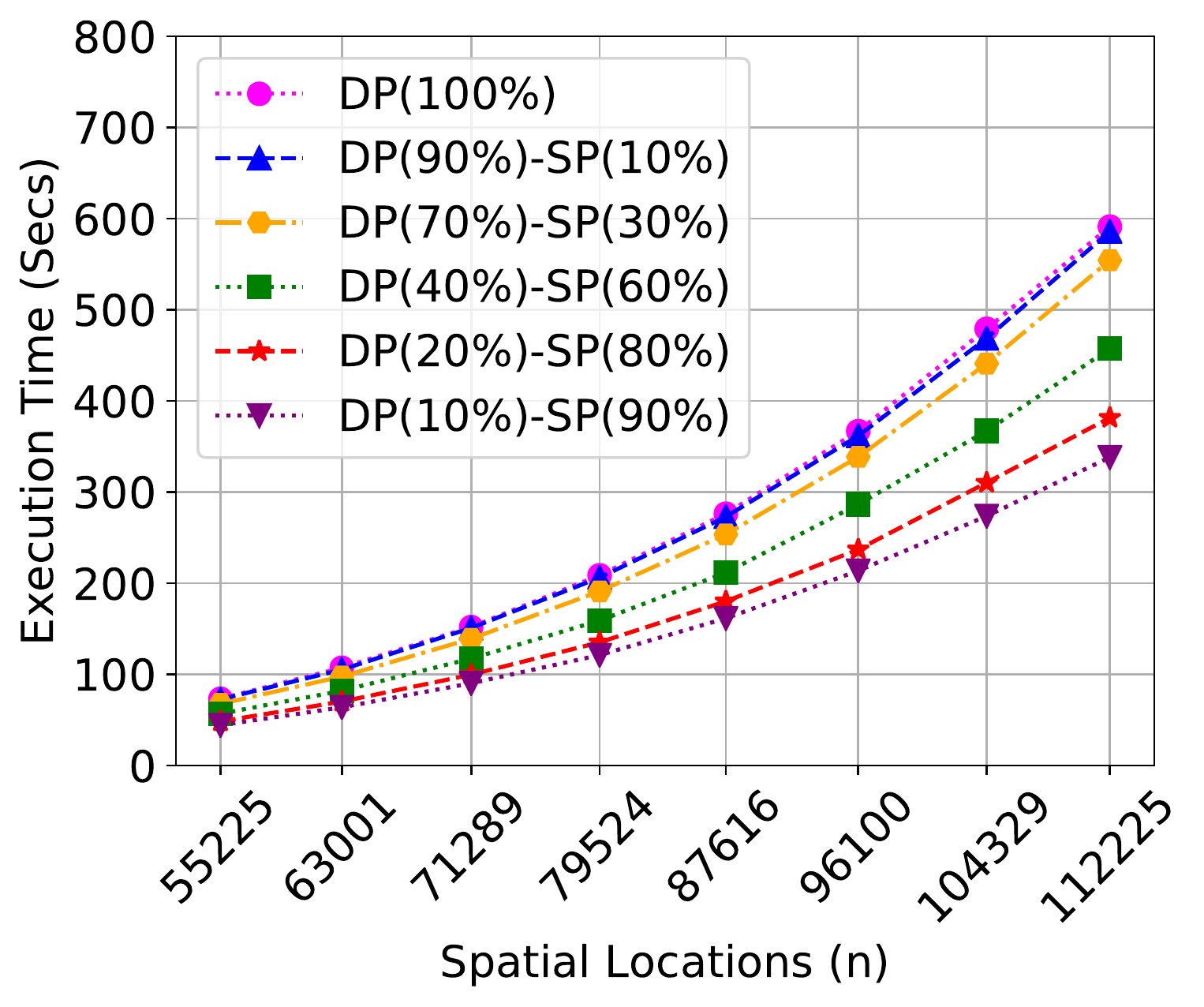}
	 	}
	 	\subfigure[Intel Skylake (56-core).]{
	 		\label{fig:time-skylake}
	 	\includegraphics[width=0.65\linewidth]{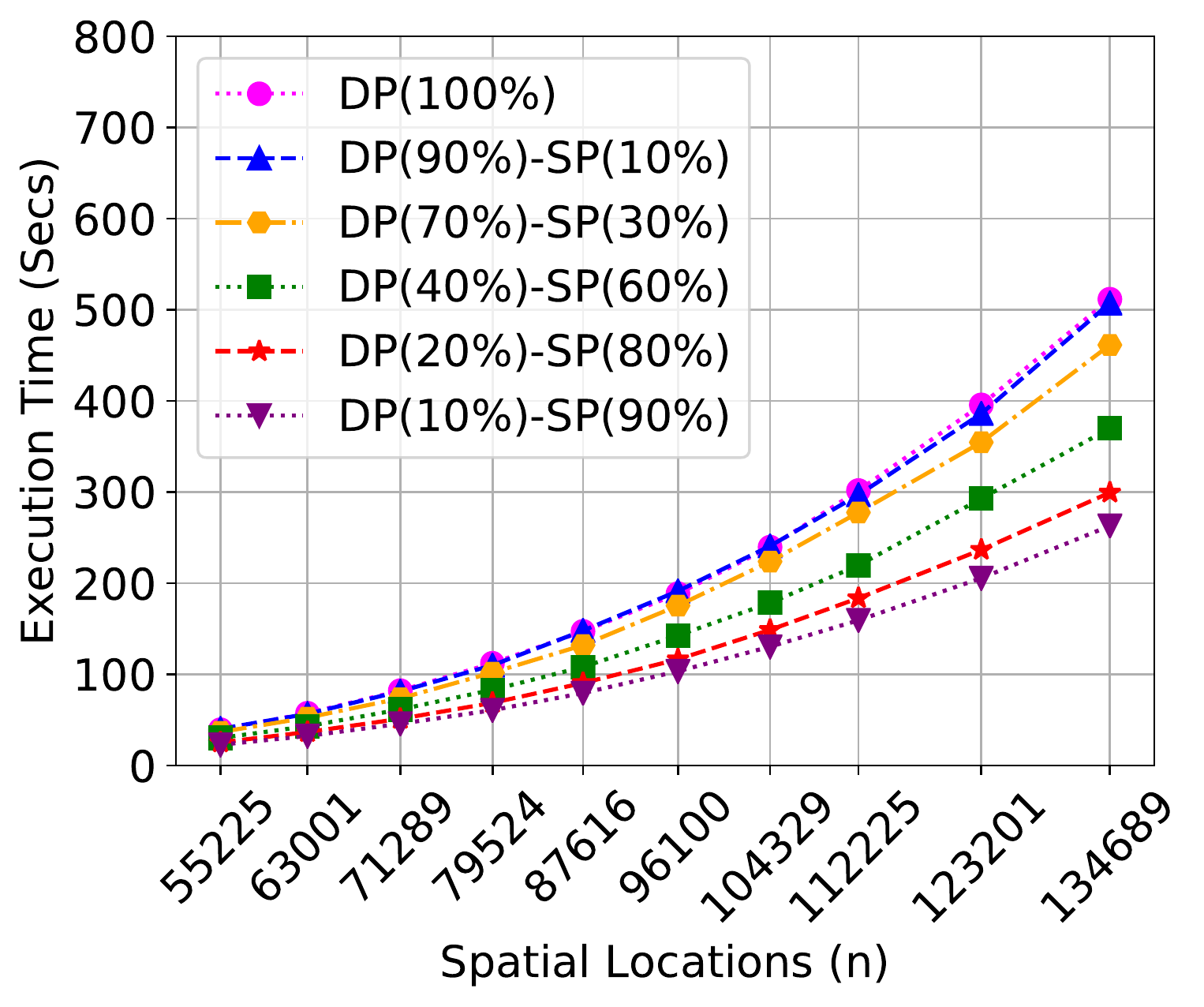}}
	 	\caption{Execution time per iteration on Intel shared-memory architectures.}
	 	\label{fig:exe-time-shared}
	 \end{figure}

	 we performed the performance analysis of the mixed-precision
	 MLE method on two recent Intel shared-memory architectures, Intel Haswell
	 and Intel Skylake processors, over different problem size up to $\sim134K$ spatial
	 locations as shown in Fig.~\ref{fig:exe-time-shared}. The x-axis represents the number of 
	 spatial locations
	 $n$, and the y-axis represents the  execution time per iteration in seconds.
	 We reported the execution time to show the performance improvement
	 accomplished by the mixed-precision method.
	
	 Fig.~\ref{fig:time-haswell} shows the average execution time on a 36-core Intel Haswell
	 processor for different mixed-precision configurations.
	 As shown, the elapsed time for the mixed-precision method with different accuracy levels outperforms
	 the {\em DP(100\%)} variant. The average obtained speedup
	 of using  {\em DP(10\%)-SP(90\%)} variant compared to  {\em DP(100\%)} across different
	 $n$ sizes is about $1.71X$. This speedup decreases when the number
	 of diagonal {\em DP} tiles increases, i.e., the increase
	  of the {\em DP} diagonal tiles, and the decrease of the {\em SP} off-diagonal tiles.
	  The same observation can be made for the Intel Skylake processor in Fig.~\ref{fig:time-skylake}
	 where the average speedup of  {\em DP(10\%)-SP(90\%)} variant compared to the {\em DP(100\%)}
	 is about $1.84X$. In general, it is utmost importance to tune the tile size $nb$
	 for achieving high performance when running in mixed-precision or full accuracy mode.
	 For both processors, we have used $nb=960$ to achieve the best performance.

	Obviously, from the two figures, the average execution time has an inverse proportion with the
	obtained accuracy at the end. This behavior is expected since using more single-precision tiles
	should reduce the execution time and at same the time should decrease the estimation accuracy.
	In section~\ref{sec:estimation}, we performed a set of  experiments on synthetic datasets
	and real datasets to show the estimation and the prediction accuracy difference between different mixed-precision variants
	and the double-precision method.

%

	\subsubsection{Heterogeneous Shared-Memory Architectures}
	we estimated the average execution time of the likelihood evaluation function, using
	shared-memory systems equipped with GPU accelerators. 
	The overall execution time on CPU/GPU systems involving both the
	computation time and the data movement time from the CPU memory
	to the GPU memory or vice versa. Thus, in our study, we estimated the total
	execution time and the required data movement for both the double-precision
	and mixed-precision methods. 

\begin{figure*}[!ht]
		\scriptsize
		\centering
		\subfigure[Intel Broadwell w/ K80 GPU.]{
			\label{fig:k80-gpu}
		\includegraphics[width=0.3\linewidth]{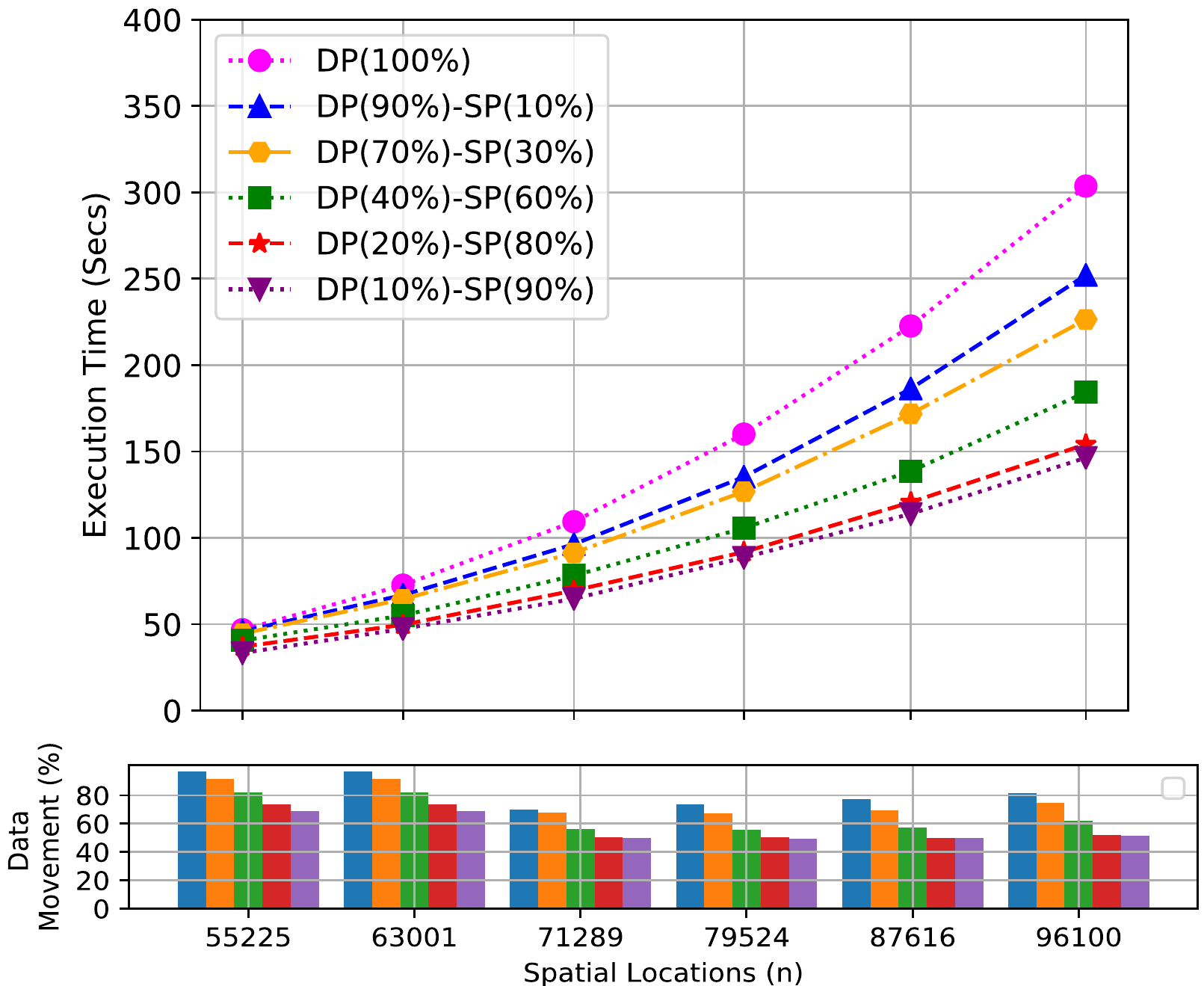}}\hspace{0.1in}
				\subfigure[Intel Haswell w/ P100 GPU.]{
			\label{fig:p100-gpu}
		\includegraphics[width=0.3\linewidth]{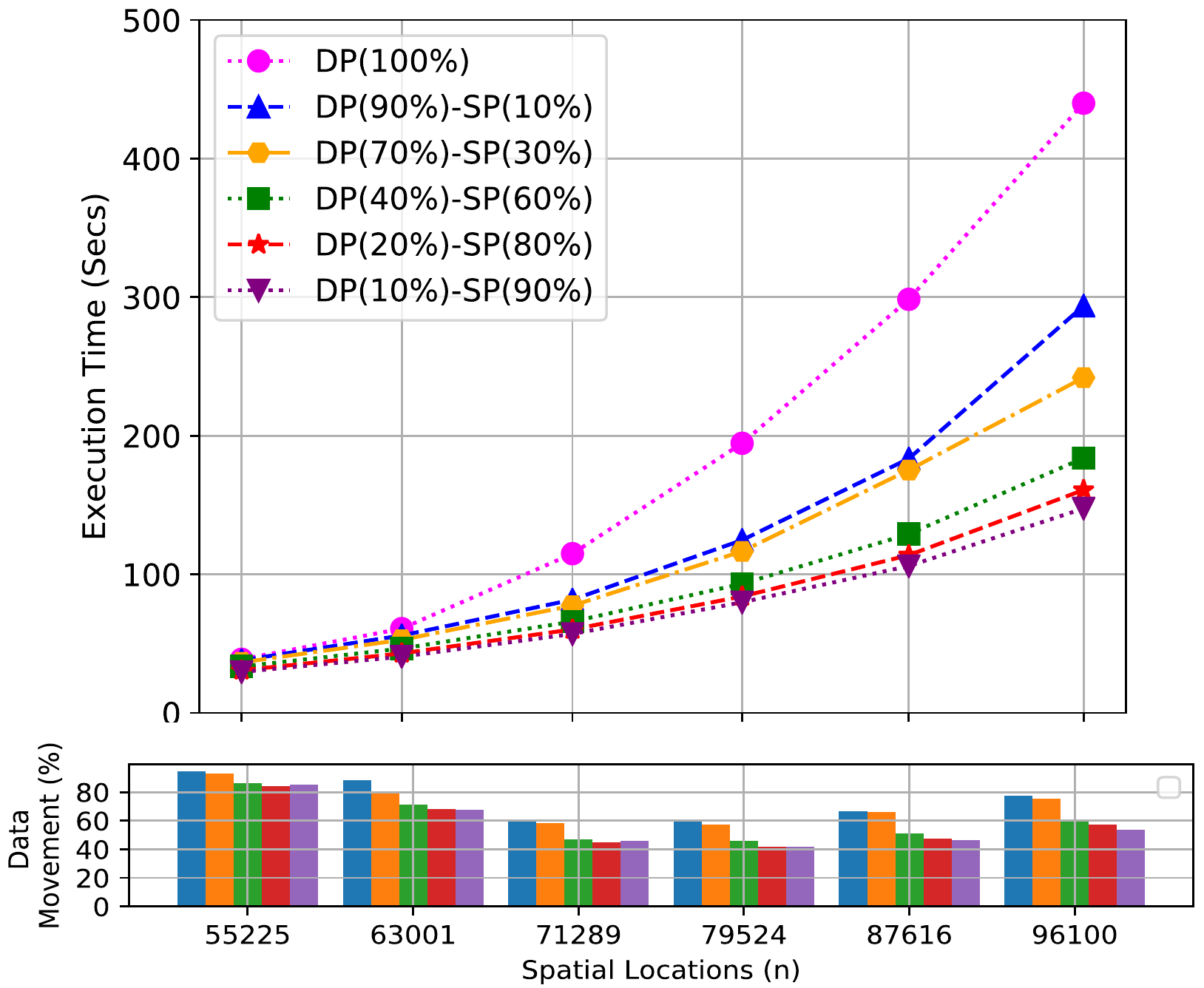}}\hspace{0.1in}
			\subfigure[Intel Skylake w/ V100 GPU.]{
			\label{fig:v100-gpu}
			\includegraphics[width=0.288\linewidth]{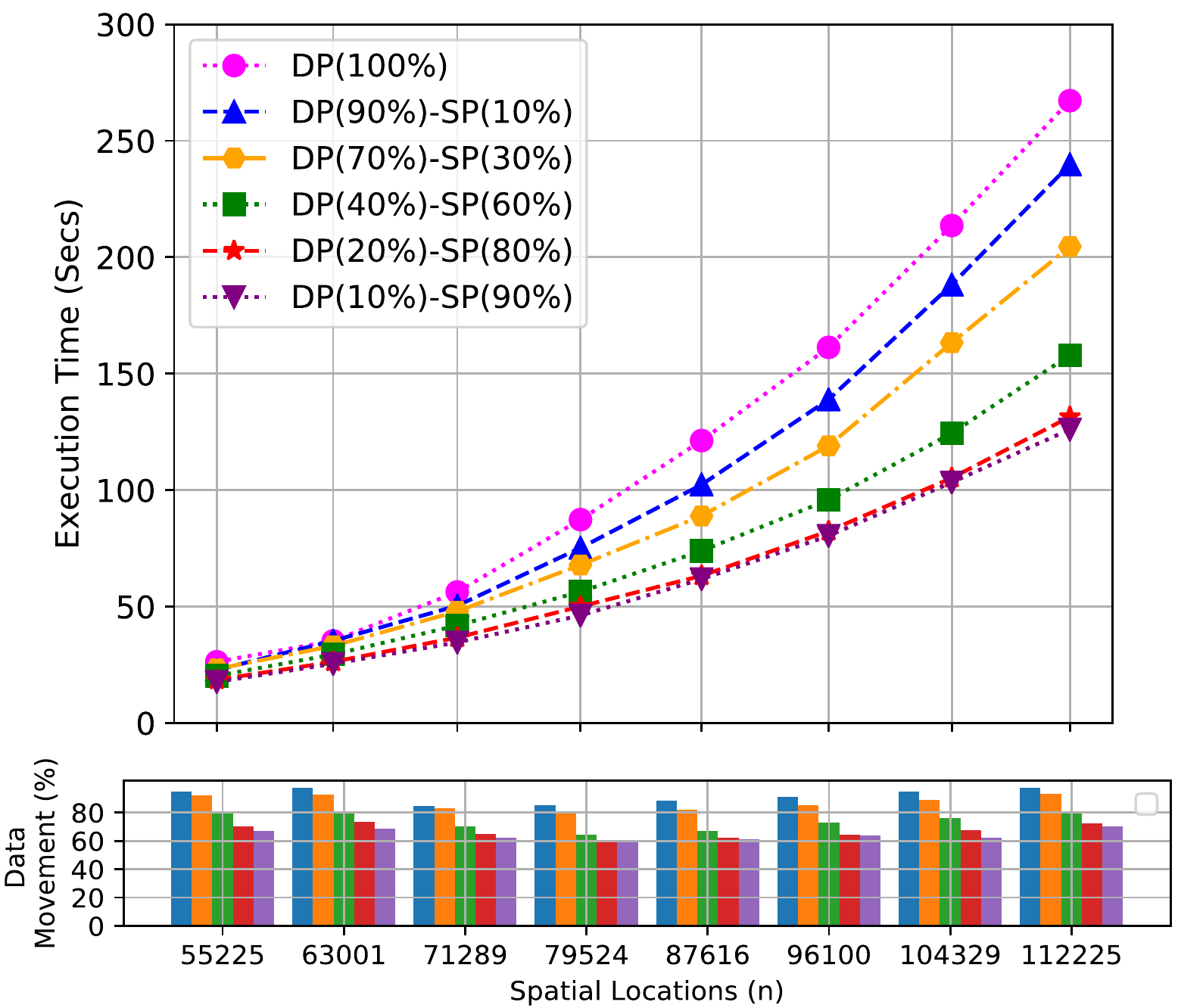}
		}
		
		\caption{Time and data movement cost per iteration on shared-memory architectures equipped with GPUs.}
		\label{fig:gpus}
	\end{figure*}

	Fig.~\ref{fig:k80-gpu} shows the total execution time, using an Intel Broadwell
	equipped with a Tesla K80 Kepler GPU. The average speedup that we obtained
	by the mixed-precision variant  {\em DP(10\%)-SP(90\%) } was $1.74X$ with upto
	$\sim96K$ spatial locations.  The cost of data movement
	for each mixed-precision variant and for each $n$, in comparison with,
	the {\em DP} arithmetic method, are also shown in Fig.~\ref{fig:k80-gpu}.
	The {\em DP(10\%)-SP(90\%) } can reduce the data movement amount
	upto $50\%$ compared to the {\em DP(100\%)} variant. For example,
	with $n=96K$ the data movement cost of the {\em DP(10\%)-SP(90\%) }  variant is
	$1562.07$ GB  while the {\em DP(100\%)} variant requires about $3032.73$ GB. 
	In Fig.~\ref{fig:p100-gpu}, 
   the obtained speedup  by the {\em DP(10\%)-SP(90\%) }  variant is $2.18X$ in average 
	compared to the {\em DP(100\%)} variant using an Intel Haswell equipped with  
	a Tesla P100 Pascal GPU. We found that the difference of data movement between the two variants could reach up to $40\%$.	
In Fig.~\ref{fig:v100-gpu}, 
	the speedup reached $1.82X$  whereas the data movement were reduced to as much as $60\%$ of the {\em DP(100\%)} variant	using an Intel Skylake equipped with a Tesla V100 Volta GPU.

In practice, StarPU  moves data around much more than expected, due to its aggressive perfecting strategy. Reducing the data transfer amounts between the GPU memory and CPU
	 memory reduces the	data communication overhead and increases the overall
	 obtained speedup which can reach more than $2X$ in some cases.

\begin{figure*}[!ht]
		\centering
			\subfigure[64 Nodes.]{
			\label{fig:64-nodes}
		\includegraphics[width=0.3\linewidth]{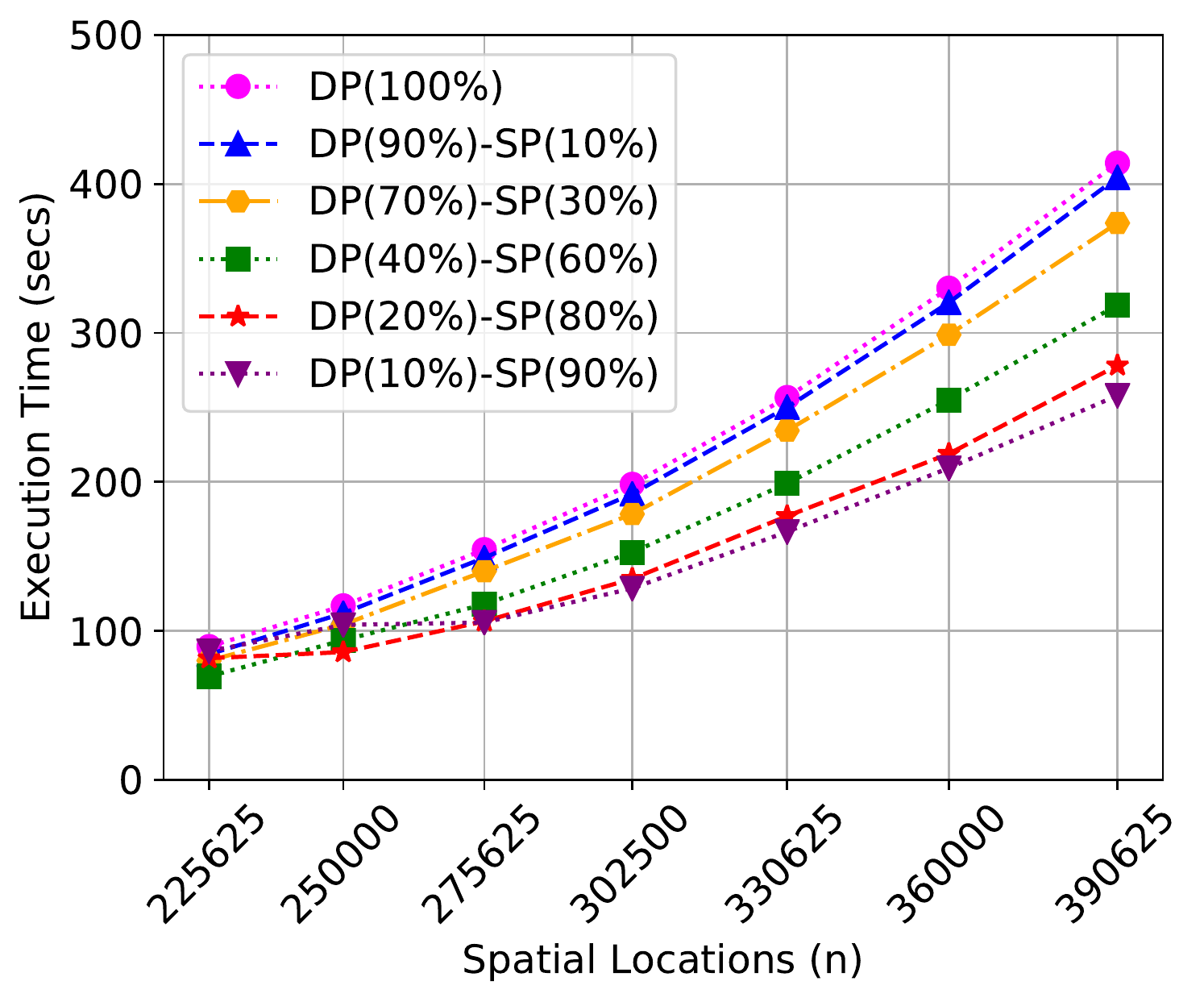}}
				\subfigure[128 Nodes.]{
			\label{fig:128-nodes}
		\includegraphics[width=0.3\linewidth]{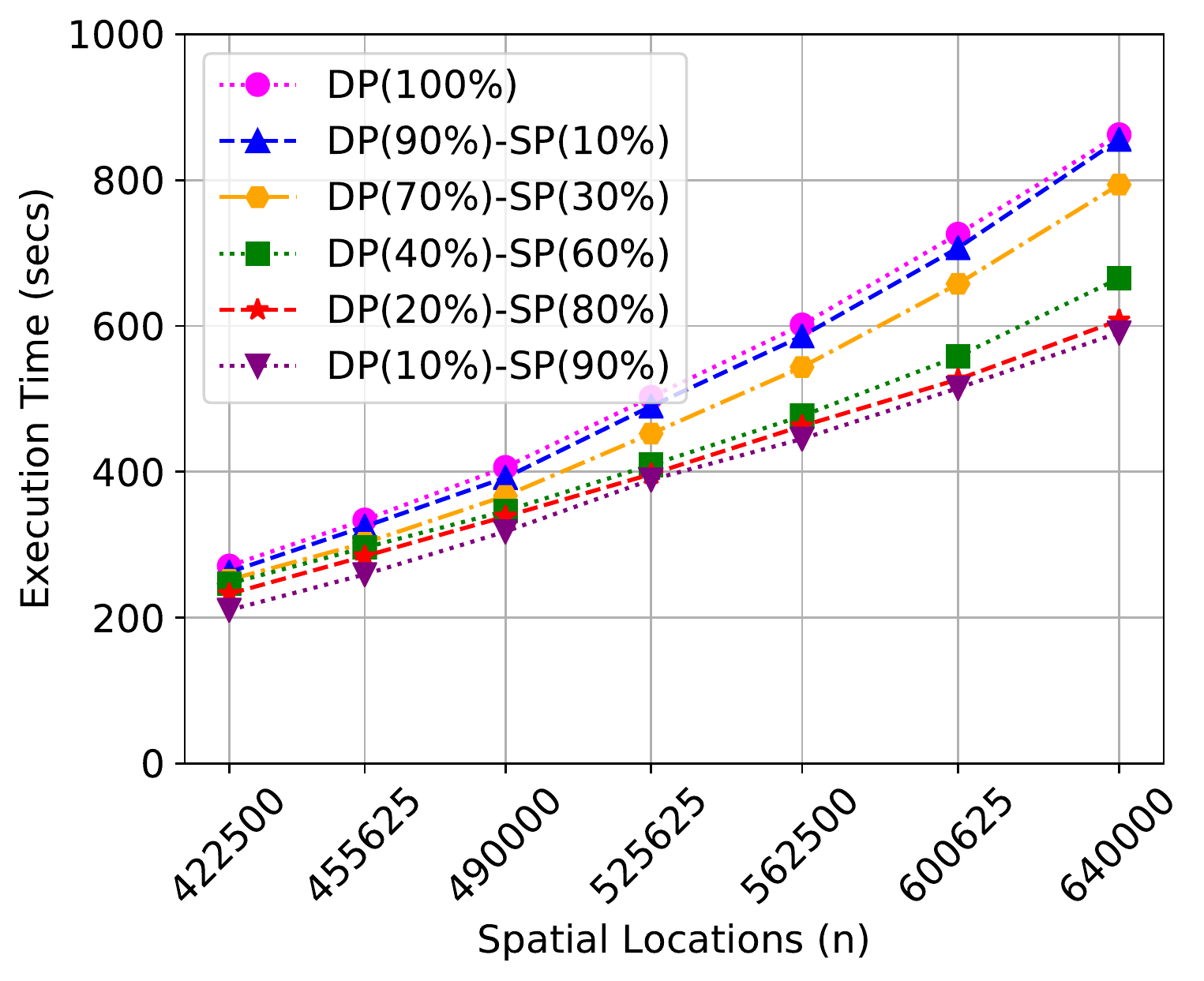}}
		\subfigure[Scalability with different \# of nodes.]{
			\label{fig:scala}
		\includegraphics[width=0.3\linewidth]{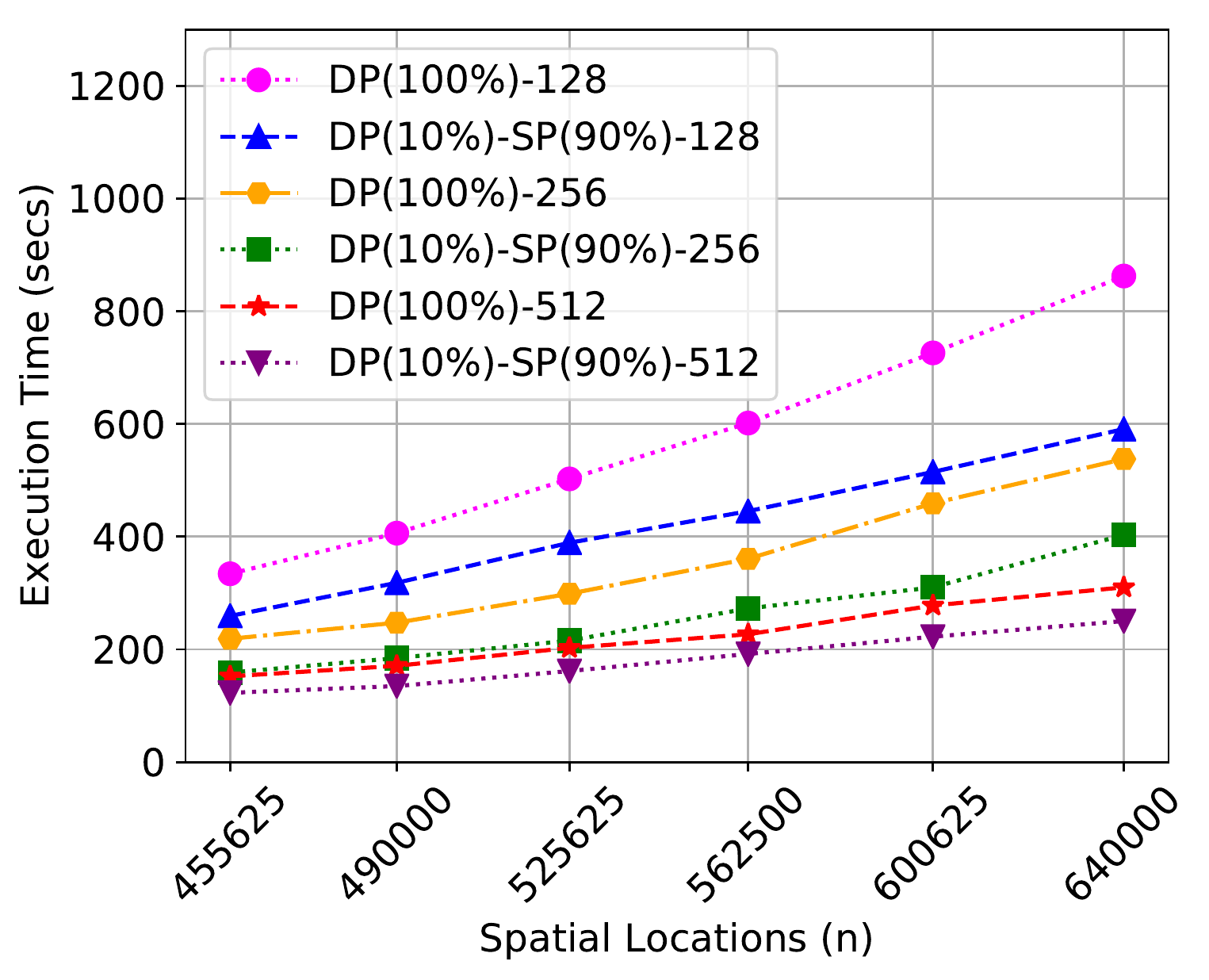}}
		\caption{Execution time per iteration on different number of nodes on Shaheen-II (Cray XC40).}
		\label{fig:time-shaheen}
	\end{figure*}

	\subsubsection{Distributed-Memory Systems}
	
	we present the performance analysis of the proposed mixed-precision computation
	on the distributed-memory Shaheen-II Cray XC40 system using a different number of
	nodes, 64, 128, 256, and 512 (i.e., up to 16384 cores). Fig.~\ref{fig:64-nodes} and Fig.~\ref{fig:128-nodes} show the execution time
of different mixed-precision variant compare to the double-precision method.
The figures show a speedup upto $1.61X$ and $1.45X$ when using 64 and 128 nodes. Moreover, Fig.~\ref{fig:scala} shows the scalability with different number of nodes, 128, 256, and 512.
 As also shown in Fig.~\ref{fig:64-nodes} and Fig.~\ref{fig:128-nodes},  the mixed-precision method shows a linear scaling behavior with different numbers of nodes, and the gained up
  speedup for 256, and 512 nodes can reach $1.48X$ and $1.27X$, using the mixed-precision method.

	\subsection{MLE Statistical Parameters Estimation/Predictions Accuracy}
	\label{sec:estimation}
	
	The experiments in the previous subsection were geared towards the performance
	evaluation across different hardware architectures. Here, we aim to evaluate
	the effectiveness of our proposed mixed-precision method for the MLE calculations
	in terms of parameter vector estimation accuracy and prediction error
  compared to the DP arithmetic method. 
  To do so, we used both synthetic and real datasets to evaluate the accuracy
	and to show the effectiveness of the mixed-precision method.

	\subsubsection{Synthetic Datasets}
		
	\begin{figure*}[!ht]
		\centering
		\subfigure[Weak correlation ($\theta_2 = 0.03$).]{
		\label{fig:weak}
		\includegraphics[width=0.3\linewidth]{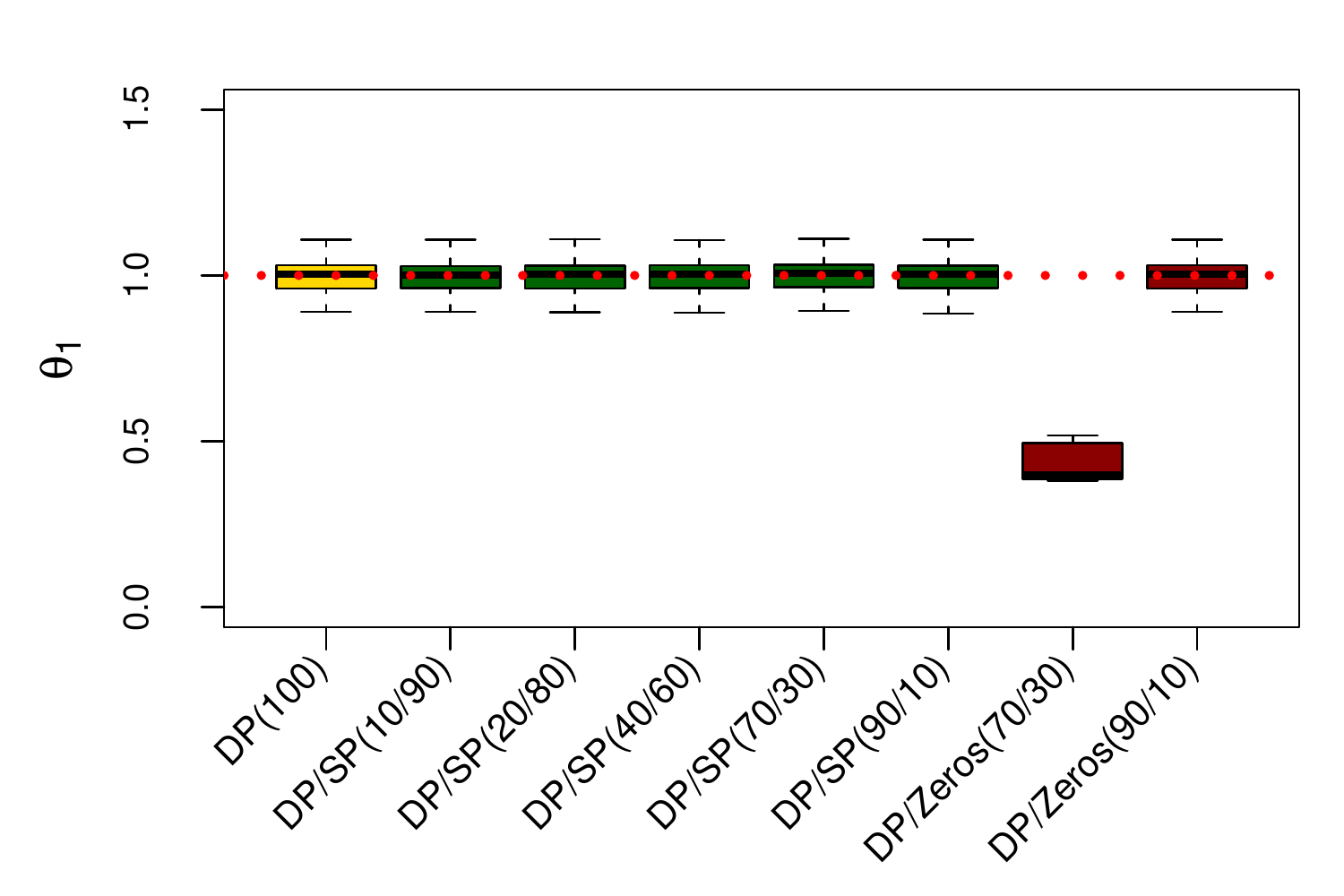}
		\includegraphics[width=0.3\linewidth]{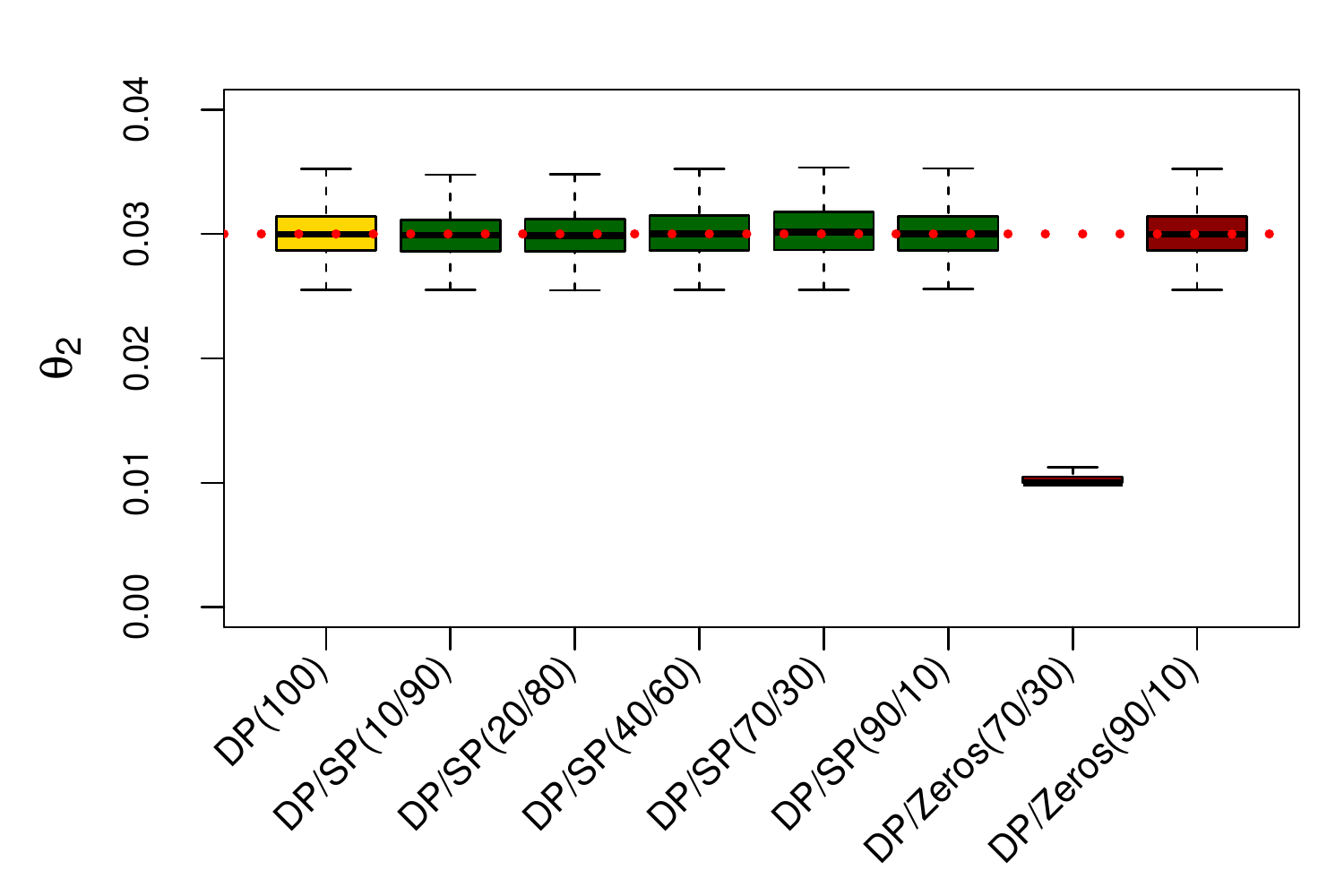}
		\includegraphics[width=0.3\linewidth]{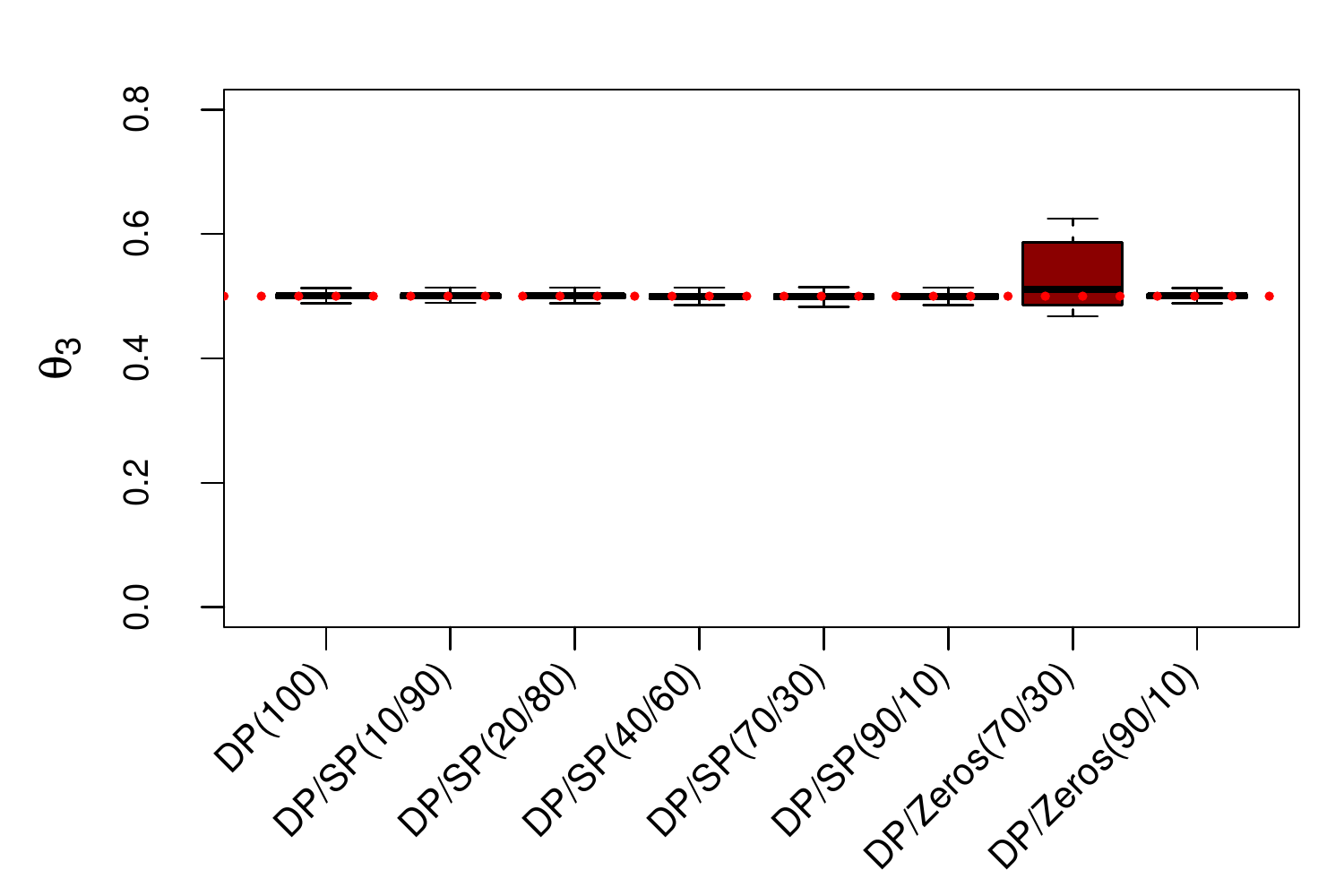}
		}

	     \vspace{-0.6\baselineskip}

		\subfigure[Medium correlation ($\theta_2 = 0.10$).]{
		\label{fig:medium}
		\includegraphics[width=0.3\linewidth]{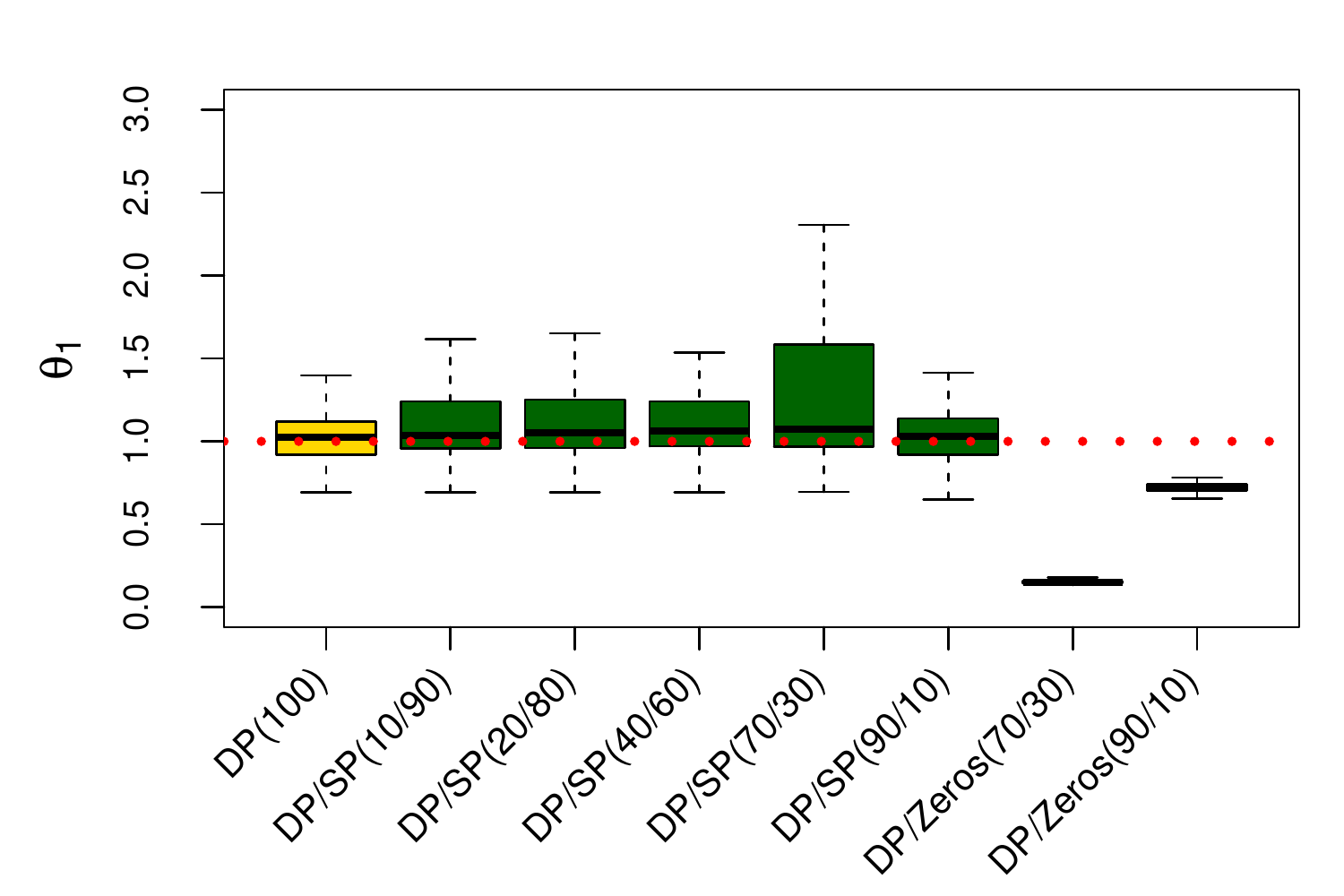}
		\includegraphics[width=0.3\linewidth]{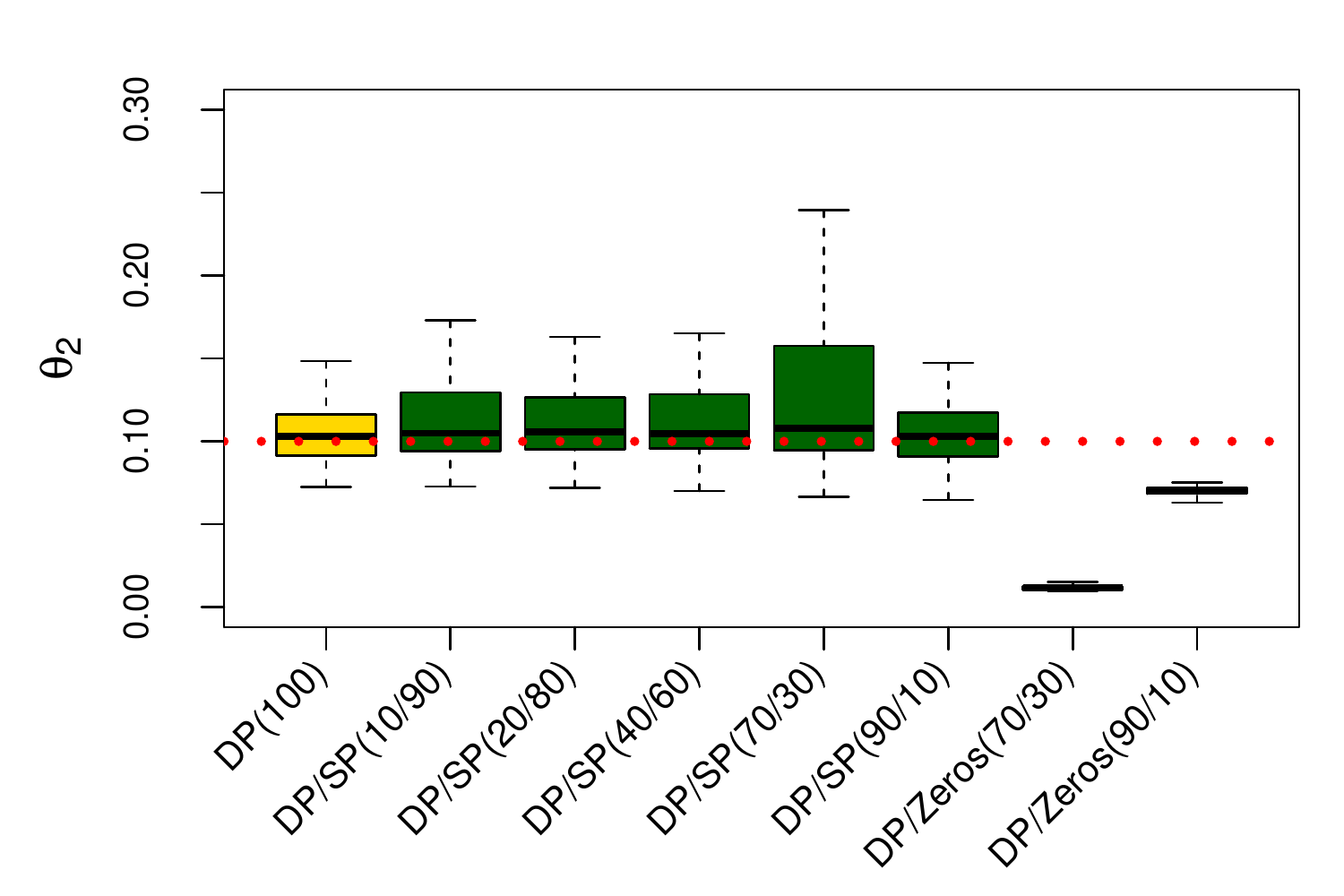}
		\includegraphics[width=0.3\linewidth]{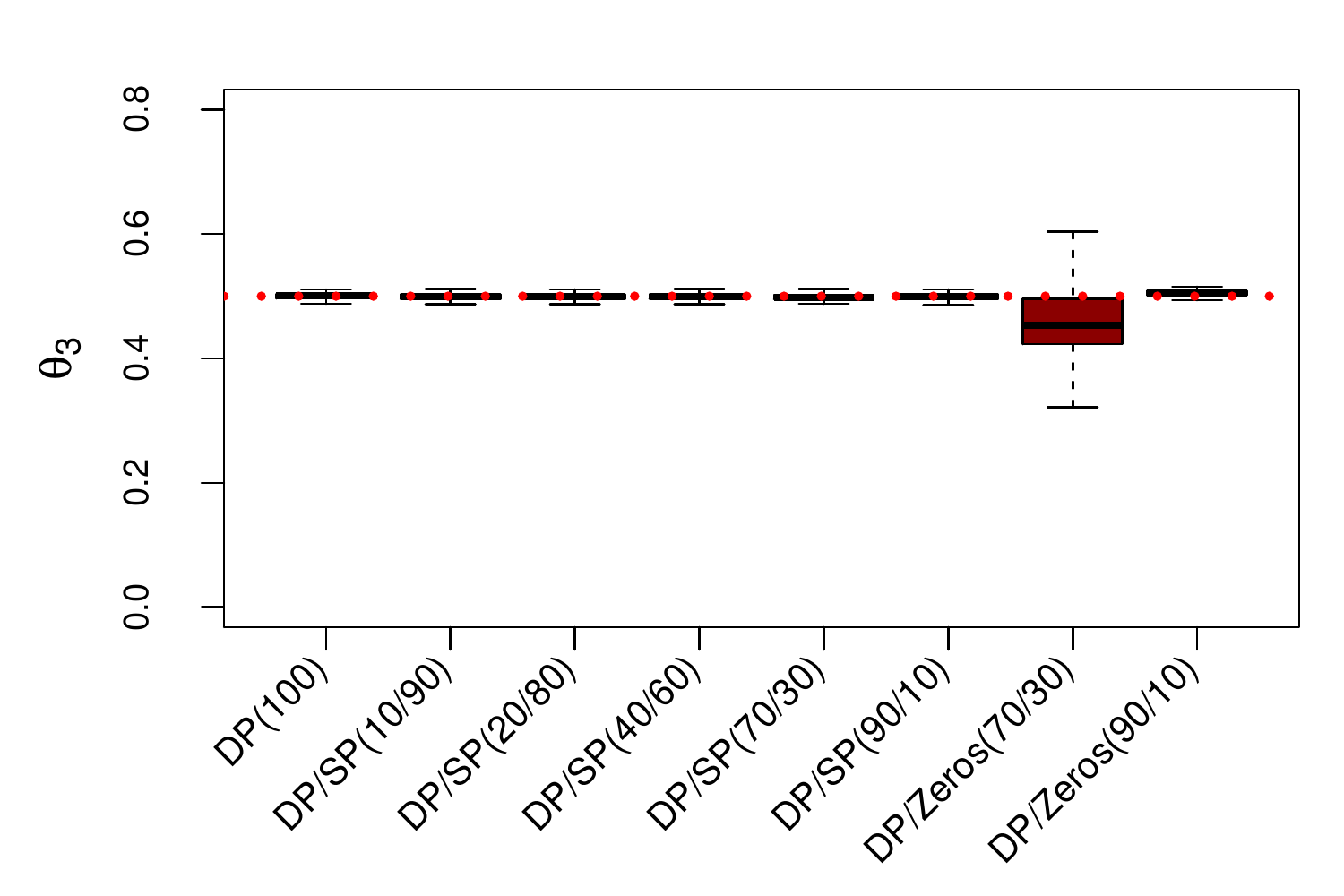}
		
		}
	
		     \vspace{-0.6\baselineskip}

		\subfigure[Strong correlation ($\theta_2 = 0.30$).]{
		\label{fig:strong}
		\includegraphics[width=0.3\linewidth]{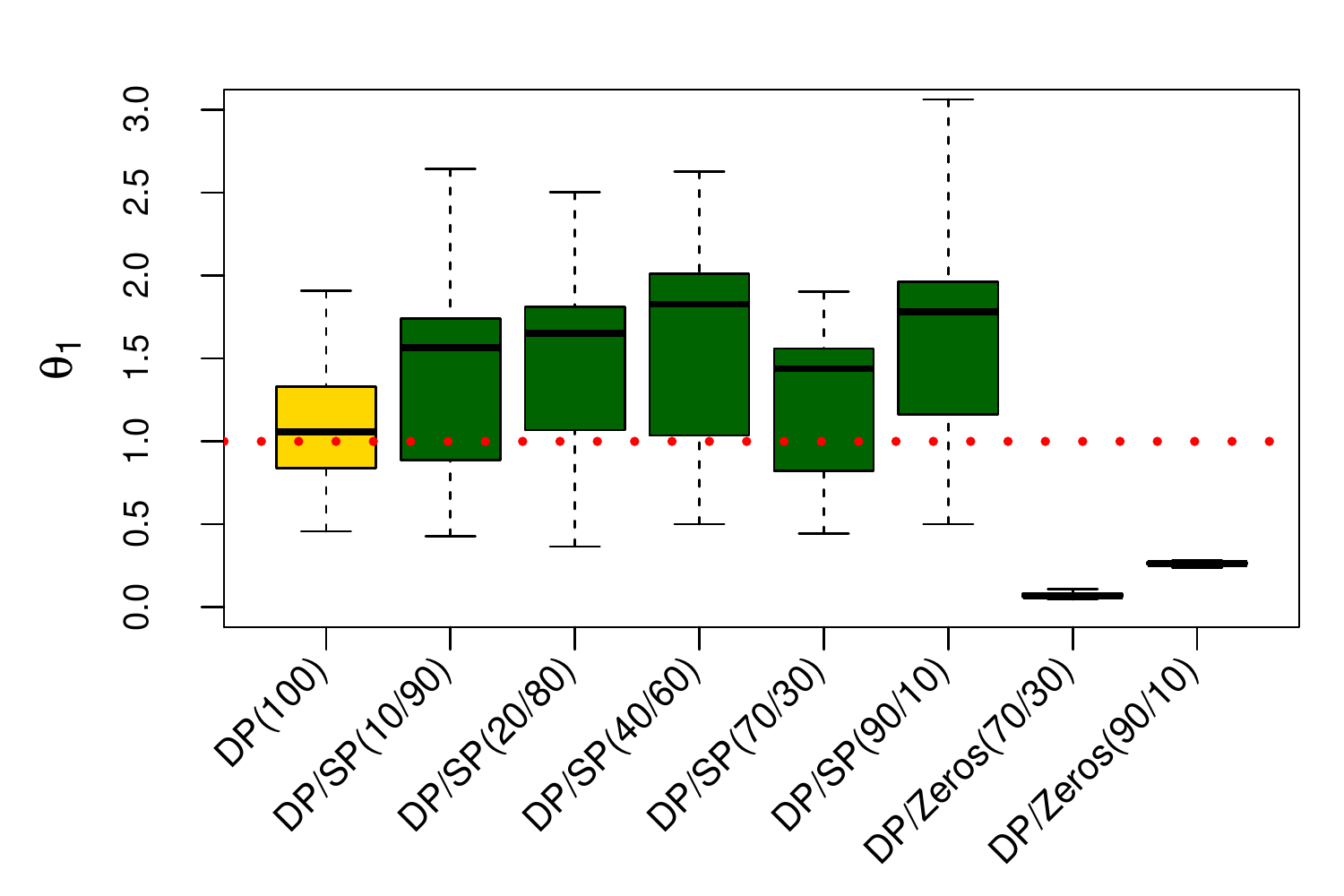}
		\includegraphics[width=0.3\linewidth]{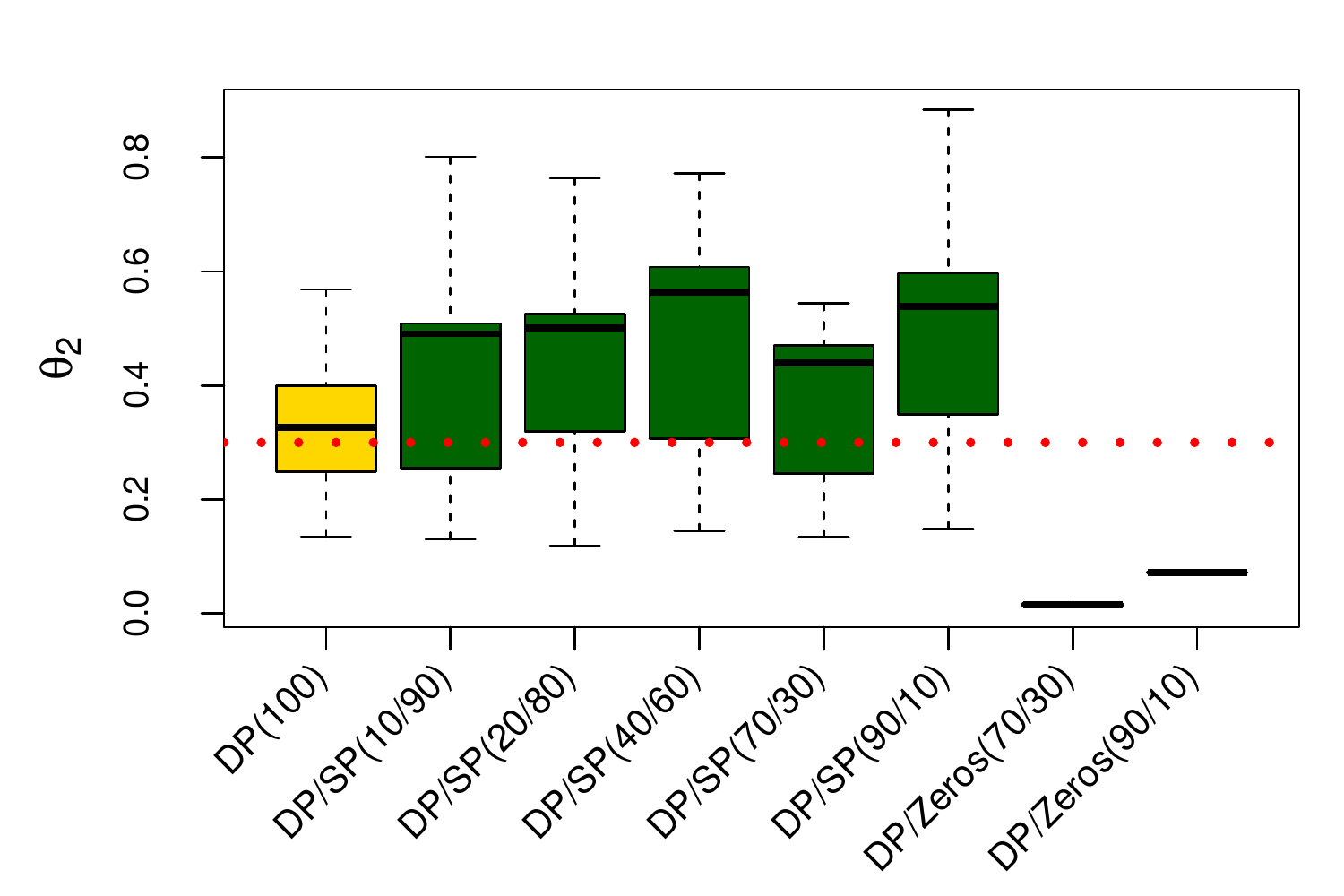}
		\includegraphics[width=0.3\linewidth]{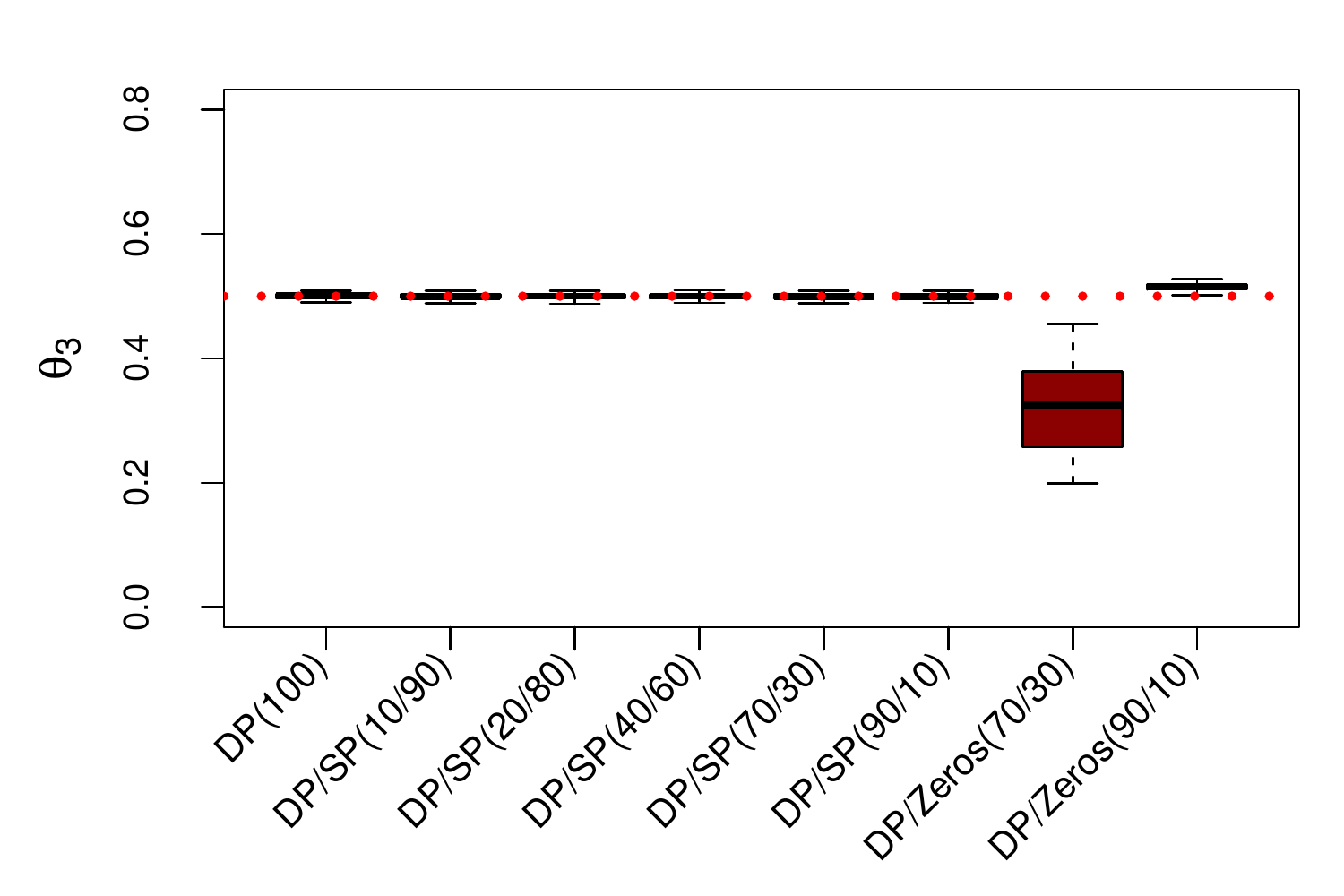}
	
		}
				\caption{{Parameters estimation boxplots of 40K synthetic datasets using {\em DP},  {\em DP(x\%)}-{\em SP(y\%)}, and {\em DST} MLE algorithm. True values of the parameters  are represented by red dotted line when estimating the parameter vector.} 
		}
		\label{fig:estimates}
	\end{figure*}
	
	The MLE operation involves estimating the model parameter vector
	$\widehat{\boldsymbol \theta} = (\theta_1$, $\theta_2$, $\theta_3$) for the underlying data and uses
	these estimated  parameters to predict missing values at other known
	spatial locations. We performed a set of experiments to evaluate
	the accuracy of our proposed mixed-precision method for the MLE calculations,
	based on Monte Carlo simulations.  The simulations required the availability of
	synthetic spatial datasets with different properties to cover different cases
	that are expected to exist in real data. For instance,  $\theta_2$, which represents
	the correlation degree between the given spatial locations, could be weak,
	medium, or strong. Thus, accuracy verification should involve datasets with different
	correlation degrees to properly evaluate the proposed MLE computation method. 
	
	Recalling section~\ref{subsec:data_generator}, \exageostat the data generator
	is used to generate a set of 2D locations with a set of  associated $\boldsymbol Z$
	measurement vectors. We produced a set of $40K$ synthetic datasets that represents the
	three correlation levels, the weak correlation ($\theta_2=0.03$), the medium ($\theta_2=0.10$),
	and the strong ($\theta_2=0.30$) correlation levels. For each case, we generated 100 different
	spatial data (i.e., locations and measurements). We used the DP
	 method, five levels for mixed-precision  method,
	i.e, {\em DP(10\%)-SP(90\%), 
	DP(20\%)-SP(80\%), DP(40\%)-SP(60\%), DP(70\%)-SP(30\%), DP(90\%)-SP(100\%)},
	 and two levels of the DST method,
	i.e.,  {\em DP(70\%)-Zero(30\%)} and  {\em DP(90\%)-Zero(10\%)}, for accuracy comparison. 
		We have ignored the SP(100\%) variant because most of the interactions are captured within the
	vicinity of the diagonal tiles. Therefore, it is critical to ensure high precision computations (i.e. double-precision)
	around the diagonal tiles. If single-precision is used instead, the covariance matrix may lose the numerical property
	of positive definiteness, and the MLE procedure cannot proceed.	We also ignores comparisons against TLR
	 because it out of the scope 	of this paper and it requires a thorough analysis,	since it may yield  poor results when nuggets (noise) are relatively small
	and observations are mostly dense~\cite{stein2014limitations}.
	Different computation methods
	were used to estimate the model parameter vector 
	in order to show that the estimated parameter vector $\widehat{\boldsymbol \theta}$ was consistent with the
	initial parameter vector $\boldsymbol {\theta_0}$ that had been used to generate the 
	spatial data.

	Fig.~\ref{fig:estimates} shows the estimation accuracy results for the 40K synthetic datasets. 
	The boxplots report the results for 	100 different datasets for each correlation case.
	The experiment involved an estimation of the accuracy of the model parameters
	for weak, medium, or strong correlations (Fig.~\ref{fig:weak},
	Fig.~\ref{fig:medium}, and Fig.~\ref{fig:strong}, respectively) 
   The Prediction Mean Square Error (PMSE) is also shown for each correlation case.
	
	As shown  in Fig.~\ref{fig:weak},
	weak correlated data, i.e., $\theta_2=0.03$, required a minimum 
	number of diagonal full-precision tiles, i.e., {\em DP(10\%)-SP(90\%)}, to
	 correctly estimate	the parameter vector. At the same time, the DST method required
	  at least 90\% of the tiles to properly estimate
	the parameter vector. 	Fig.~\ref{fig:medium} shows  the estimated parameter vector accuracy with  medium
	 correlated data, i.e.,  $\theta_2=0.10$. Compared to the weakly correlated, the mixed-precision
	method is less accurate but is still close to the correct values, whereas the  DST method fails to estimate
	the accurate parameters with both used variants. Fig.~\ref{fig:strong} is related to strongly
	correlated data, i.e.,  $\theta_2=0.30$. More accuracy is lost when a mixed-precision is used
	but it is still acceptable, compared with the DST method. These results were expected, because of the tight fit between the spatial data
	correlation strength and the amount of numeric loss in both
	mixed-precision and DST methods.  A higher correlation strength requires  a more accurate representation of the spatial covariance matrix.


					\begin{figure*}[!ht]

		\centering
			\subfigure[Weak correlation ($\theta_2 = 0.03$).]{
				\includegraphics[width=0.3\linewidth]{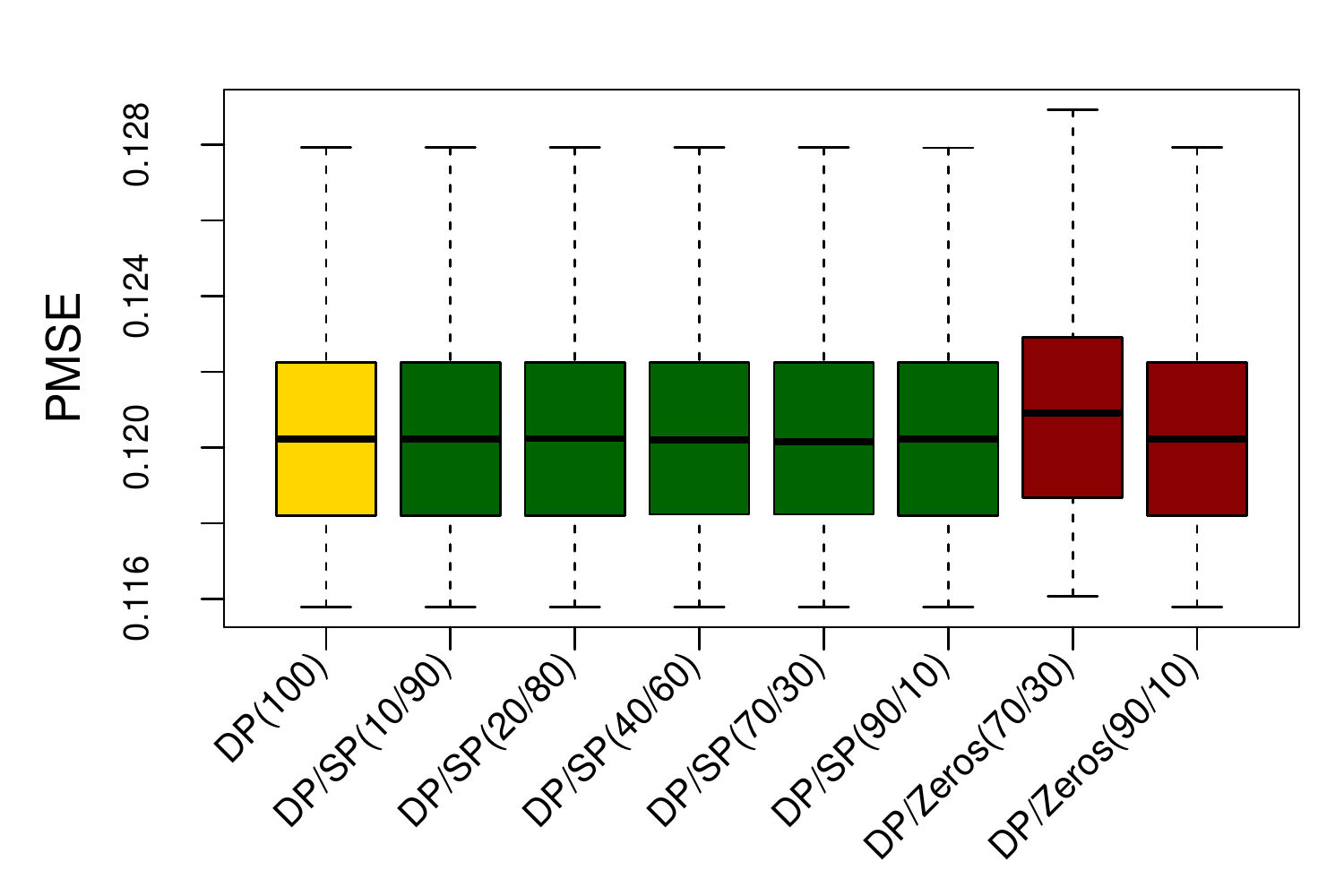}
				}
					\subfigure[Medium correlation ($\theta_2 = 0.10$).]{
				\includegraphics[width=0.3\linewidth]{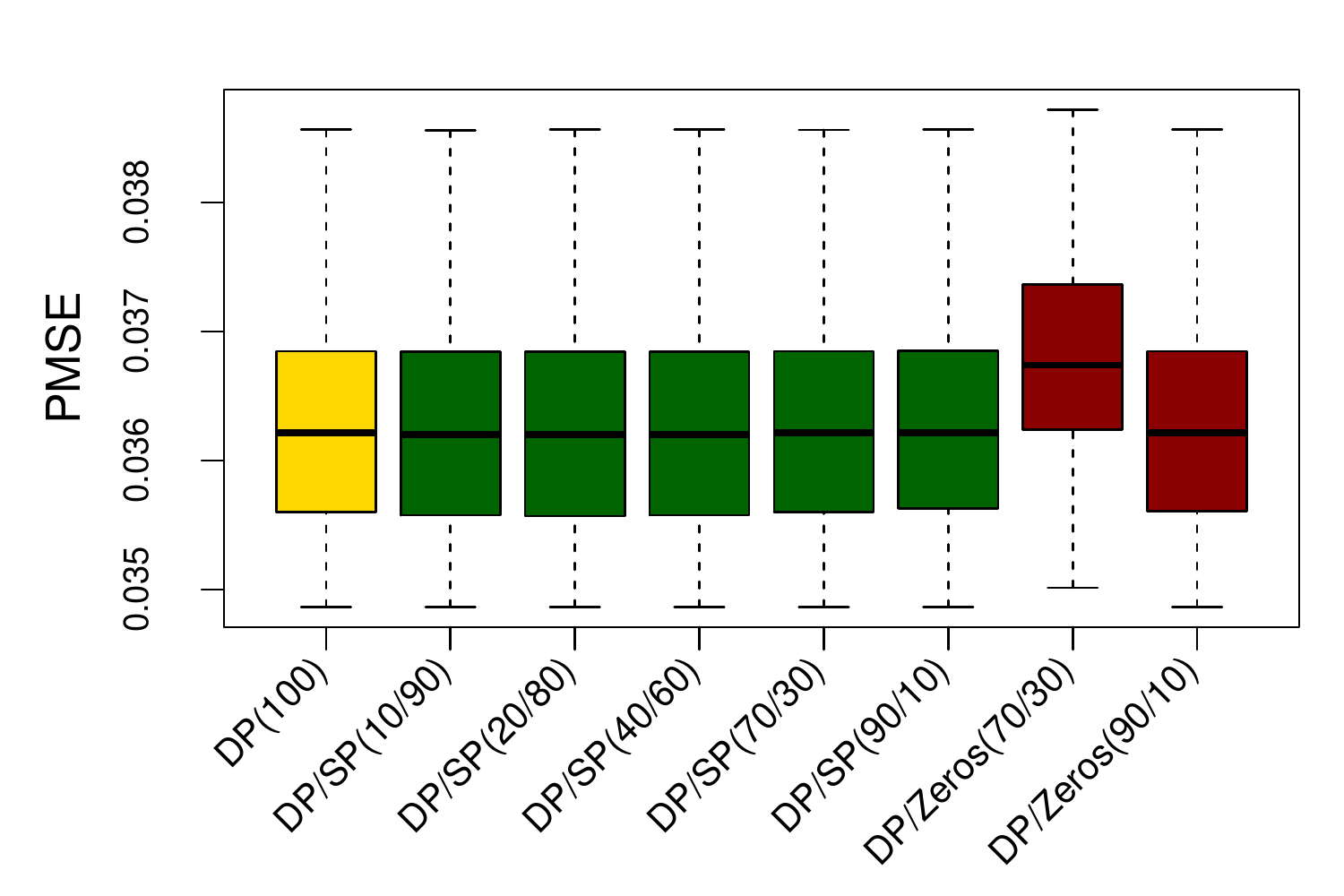}
				}
					\subfigure[Strong correlation ($\theta_2 = 0.30$).]{
					\includegraphics[width=0.3\linewidth]{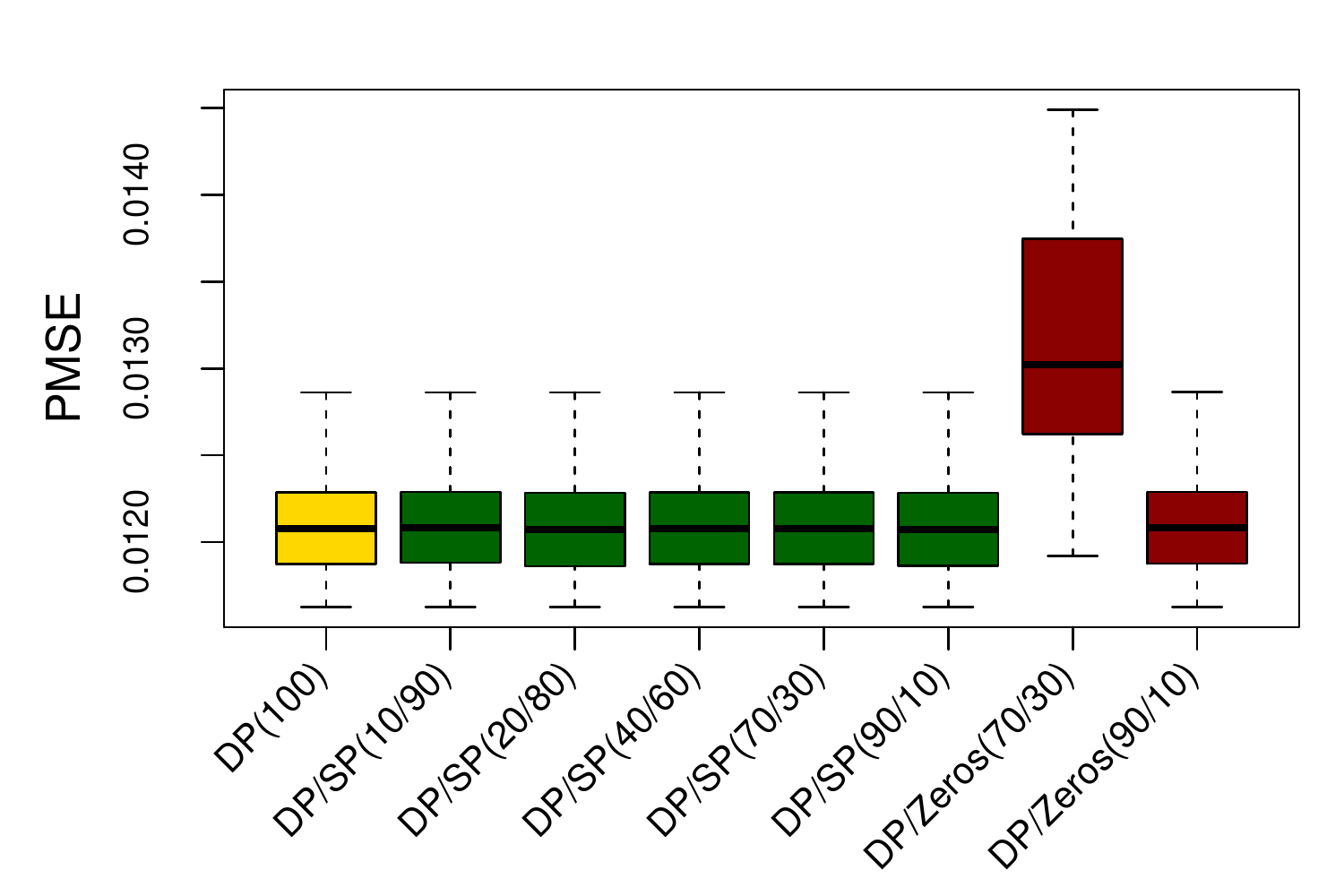}
					}
						\caption{{Prediction Mean Square Error (PMSE) boxplots using $k$-fold cross-validation technique, where $k=10$, of 40K synthetic datasets using {\em DP},  {\em DP(x\%)}-{\em SP(y\%)}, and {\em DST} MLE algorithm.} 
		}
		\label{fig:pred}
	\end{figure*}

		\begin{table*}
	\centering
		\setstretch{0.8}
	\scriptsize
	\setlength{\tabcolsep}{0.9pt}
	
	\caption{The Mat\'{e}rn covariance parameters estimation and the Prediction Mean Square Error (PMSE) using $k$-fold cross-validation technique, where $k=10$, for $4$ geographical regions of wind speed dataset.}
	\renewcommand{\arraystretch}{1.3}
	\begin{tabu}{|c|[1pt]c|c|c|c|c|c|[1pt]c|c|c|c|c|c|[1pt]c|c|c|c|c|c|[1pt]c|c|c|c|c|c|}
	\hline
	& \multicolumn{18}{c|[1pt]}{Mat\'{e}rn Covariance}  \\
	R & \multicolumn{6}{c|[1pt]}{Variance ($\theta_1$)}  &  \multicolumn{6}{c|[1pt]}{Spatial Range ($\theta_2$)}  &  \multicolumn{6}{c|[1pt]}{Smoothness ($\theta_3$)} &  \multicolumn{6}{c|}{Prediction Accuracy (PMSE)}\\

	& &\multicolumn{3}{c|[0.5pt]}{mixed-precision}  &  \multicolumn{2}{c|[1pt]}{DST} 
	& &\multicolumn{3}{c|[0.5pt]}{mixed-precision}  &  \multicolumn{2}{c|[1pt]}{DST}  
	& &\multicolumn{3}{c|[0.5pt]}{mixed-precision}  &  \multicolumn{2}{c|[1pt]}{DST} 
	& &\multicolumn{3}{c|[0.5pt]}{mixed-precision}  &  \multicolumn{2}{c|[0.5pt]}{DST} 
	\\

	& &\multicolumn{3}{c|[0.5pt]}{(DP/SP)}  &  \multicolumn{2}{c|[1pt]}{(DP/Zero)} 
	& &\multicolumn{3}{c|[0.5pt]}{(DP/SP)}  &  \multicolumn{2}{c|[1pt]}{(DP/Zero)}  
	& &\multicolumn{3}{c|[0.5pt]}{(DP/SP)}  &  \multicolumn{2}{c|[1pt]}{(DP/Zero)} 
	& &\multicolumn{3}{c|[0.5pt]}{(DP/SP)}  &  \multicolumn{2}{c|[0.5pt]}{(DP/Zero)} 
	\\

	&   \tiny $DP $ &  \tiny  $10/90$ &    \tiny  $40/60$  &   \tiny  $90/10$ &   \tiny $70/30$  &  \tiny  $90/10$
	&   \tiny $DP $ &  \tiny  $10/90$ &  \tiny  $40/60$  &   \tiny   $90/10$ &   \tiny $70/30$  & \tiny   $90/10$
	&   \tiny $DP $ &  \tiny  $10/90$ &  \tiny  $40/60$  &   \tiny   $90/10$ &   \tiny $70/30$  & \tiny   $90/10$
	&  \tiny  $DP $ &   \tiny $10/90$ &   \tiny   $40/60$  &\tiny    $90/10$ &  \tiny  $70/30$  & \tiny   $90/10$  \\[-1pt] \tabucline[1pt]{1-25}

	R1	&	8.721	&	8.695	    &	8.724	    &		8.721	&	1.115	&	8.721	   &              
			32.104	&	31.986		&	32.104	&	32.113	    &	9.990   	&	32.104	&               
			1.208	   &	1.210		&	1.208	    &		1.208	&	0.983	&	1.208	 &					   
			0.0361	&	0.0366		&	0.0366		&	0.0361	&	0.0747	&	0.0366 \\ \hline    

	R2	&	12.533	   &	12.548	   &		12.535	&		12.533	&	1.589	 &	12.533	&  
			27.603	       &	27.640	   &		27.599	&		27.603	&	9.990     &	27.603	&    
			1.270	           &	1.269	   &	    1.270	    &		1.270	    &	0.970	  &	1.270	    &                     
		 	0.0571    &	0.0574     &	0.0571   	  	&	0.0571    &	0.0912	&	0.0571 \\ \hline  

	R3	&	10.813	   &	10.870	&		10.813	    &	10.812	&	1.230	    &	10.813	&     
			19.196	       &	19.241	&		19.188		&	19.195	&	13.880	&	19.196	&      
			1.417	           &	1.417	    &		1.417		    &	1.417	    &	0.538	    &	1.417	    &      
			0.0612      	    &	0.0614    &	    0.0612		&	0.0606 	&	0.1026	&	0.0612	 \\ \hline    

	R4	&	12.441	&	12.440	   &		12.402	   &	12.439	&	2.452	   &	12.441	&
			19.733	    &	19.732	   &		19.682		&	19.733	&	9.990	   &	19.734	&
			1.119	       &	1.119	       &		1.120		    &	1.119	   &	0.875	   &	1.119	    &
			0.2098	    &	0.2100    	&	    0.2097		&	 0.2098	&	0.6290	&	0.0678\\ \hline

		\end{tabu}
	\label{tab:regions4}
	\end{table*}	
		\subsubsection{Real Dataset}

 The estimation operations requires 	several likelihood function evaluation. 
  Assuming $10^{-3}$ optimization tolerance, we estimate the
   average number of iterations that requires by each computation
    variant to convergence. Results
  showed that high correlated data required more iterations, for both
   mixed-precision and DST methods, compared with the DP
    method. Thus, in some cases,
 the total execution time of the MLE operation using the mixed-precision
  method
 exceeded the total time required by the DP arithmetic method. 
 However,  the average number of iterations decreased with fewer correlated data.

	 The PMSE boxplots for three correlation level are shown in Fig.~\ref{fig:pred}. As shown,
 for our three cases, i.e., strong, medium, and weak correlation, the mixed-precision method
 has a prediction accuracy close to the DP arithmetic method even with the lowest accurate computation variant,
 i.e., {\em DP(10\%)-SP(90\%)}. However, the DST method only performs well when representing
 90\% of the tiles in DP representation. We also found the high correlated data to be helpful, in general, to
accurately predict missing measurements, but that the accuracy decreased with fewer correlated data.

We validated the accuracy of the proposed mixed-precision method,
using the wind dataset. We  divided it into four subregions (i.e., 1, 2, 3, and 4)
to give more examples to validate our accuracy.
 Each region has approximately
	$250K$ locations.
	Table~\ref{tab:regions4} reports the complete results of the estimation
	accuracy besides the prediction error of each computation variant of
	DP, mixed-precision, and DST methods. The prediction
	was estimated using the k-fold cross-validation technique, for k=10, to
	validate
    the prediction accuracy using different synthetic dataset sizes. The total
     number of missing values equals to n=k (i.e.,
      subsample size). 

Results showed that, across all the regions,  all the mixed-precision variants
 achieved high accuracy estimation levels, equal or at least very close to the
  estimation achieved by the DP arithmetic
 method. For the DST method, only  {\em DP(90\%)/Zero(10\%) 
   correctly 
 estimated the 
 parameters of the model. Accordingly, the prediction accuracy of all mixed
  precision variants
 were close to the DP prediction accuracy, whereas only DST {\em (DP{90\%)/Zero(10\%)}
could reach the same prediction accuracy level for all the regions.
 We also observed that with highly correlated data (i.e., Region 1 and 2),
 the mixed-precision method
required a larger number of iterations to reach convergence, compared with the
 DP arithmetic method. This number
of iterations decreased with the usage of more DP tiles in the diagonal. 
With fewer correlated
data (i.e., regions 3, and 4), the number of iterations was almost the same as 
the DP arithmetic method.

%% file: text/summary.tex
\section{Conclusion and Future Work}
\label{sec:summary}

Maximum Likelihood Evaluation (MLE) can be used to build
 a statistical model of a given set of spatial locations and observations, by estimating
the more accurate model parameter values that maximize the likelihood that the given
observations come from a distribution with these parameter values.  In Geostatistics
applications, the MLE operation involves  building
a covariance matrix which requires $O(n^3)$ floating point operators and an
$O(n^2)$ memory space to be handled in dense format. 
The Cholesky factorization is the most time-consuming operation in MLE computation.
Thus, reducing the complexity of performing it is a necessity to speedup the whole operation especially in large-scale executions.
This paper highlights a novel mixed-precision approach for the Cholesky factorization
algorithm. The application covariance matrix is built in order for double-precision and single-precision arithmetics to be applied to diagonal tiles and off-diagonal-tiles, respectively.
 The new implementation provided up to 1.6X performance speedup
on massively parallel architectures while maintaining the accuracy necessary for modeling and prediction.

In the paper, we propose an empirical approach and ensure numerical accuracy since the computed ratio of DP/SP is application-dependent. In future work, a more systematic approach can take into account the  distance between locations and switch to lower precision beyond a certain distance threshold.  We also
plan to extend the proposed mixed-precision approach to have three precision layers,
i.e., half-precision, single-precision, and double-precision. In this case, we will gain
more speedup by ignoring the accuracy in the very far off-diagonal tiles and, hopefully, keep the required accuracy.

%% file: text/ack.tex
\textbf{Acknowledgement.} The authors would like to thank 
NVIDIA Inc., Cray Inc., and Intel Corp., the Cray Center
of Excellence and Intel Parallel Computing Center
awarded to the Extreme Computing Research Center (ECRC) at KAUST.
For computer time, this research used GPU-based systems as well as Shaheen supercomputer hosted 
at the Supercomputing Laboratory at King Abdullah University 
of Science and Technology (KAUST).